\documentclass[
aps,
reprint,
superscriptaddress,
longbibliography,
nofootinbib
]{revtex4-2}

\usepackage[T1]{fontenc}
\usepackage[utf8]{inputenc}
\usepackage{amsmath,amssymb,amsfonts,amsthm,bm}
\usepackage{physics}
\usepackage{bbm}
\usepackage{slashed}
\usepackage{graphicx}
\usepackage[dvipsnames]{xcolor}
\usepackage{tikz}
\usepackage{tabularx}
\usepackage{multirow}
\usepackage{array}
\usepackage{diagbox}
\usepackage{multibib}
\newcites{supp}{References}
\usepackage[colorlinks=true,citecolor=blue,linkcolor=blue,urlcolor=blue]{hyperref}

\definecolor{mycolor}{rgb}{0.8, 0.4, 0.4}

\DeclareMathOperator*{\argmax}{arg\,max}

\begin{document}
	
	\title{
		Hayden--Preskill recovery at finite temperature on a quantum processor: dynamics and initial state from the SYK model
	}
	
	\author{Jeongho Bang}
	\email{jbang@yonsei.ac.kr}
	\affiliation{Institute for Convergence Research and Education in Advanced Technology, Yonsei University, Seoul 03722, Republic of Korea}
	\affiliation{Department of Quantum Information, Yonsei University, Incheon 21983, Republic of Korea}
	
	\author{Moongul Byun}
	\email{moongulbyun@gm.gist.ac.kr}
	\affiliation{Department of Physics and Photon Science, Gwangju Institute of Science and Technology,\\123 Cheomdan-gwagiro, Gwangju 61005, Republic of Korea}
	
	\author{Kyoungho Cho}
	\email{khcho23@yonsei.ac.kr}
	\affiliation{Institute for Convergence Research and Education in Advanced Technology, Yonsei University, Seoul 03722, Republic of Korea}
	\affiliation{Department of Statistics and Data Science, Yonsei University, Seoul 03722, Republic of Korea}
	
	\author{Keun-Young Kim}
	\email{fortoe@gist.ac.kr}
	\affiliation{Department of Physics and Photon Science, Gwangju Institute of Science and Technology,\\123 Cheomdan-gwagiro, Gwangju 61005, Republic of Korea}
	
	\author{Hyeonsoo Lee}
	\email{leeh.soo@gm.gist.ac.kr}
	\affiliation{Department of Physics and Photon Science, Gwangju Institute of Science and Technology,\\123 Cheomdan-gwagiro, Gwangju 61005, Republic of Korea}
	
	\begin{abstract}
		In the original Hayden--Preskill recovery, the post-injection scrambler and initial state are {\it not related}.
		We extend this setup in two ways: by using a SWAP gate so that the scrambler and initial state are {\it related}, and by considering recovery at {\it finite} temperature.
		For this modified protocol, we show that the information is successfully recovered in the sense that the postselection probability is non-negligible and the conditional fidelity is large.
		We find that both the postselection probability and the conditional fidelity are proportional to temperature, reflecting the reduced entanglement of the initial state at lower temperatures.
		We also derive their late-time analytic estimates under the assumption of uniform operator spreading and show that they agree well with the numerical results.
		This demonstrates that strong scrambling is important for successful information recovery.
		Implementing the protocol on an IBM superconducting processor using a binary sparse SYK Hamiltonian with $N = 8$ Majoranas, we observe that the data retain the qualitative recovery   dynamics and that a SWAP-based error-mitigation scheme improves both the postselection probability and the conditional fidelity.
	\end{abstract}
	\maketitle
	

	\section{Introduction and summary}
	The black hole information problem is one of the most fundamental problems in quantum gravity.
	Hawking's semiclassical calculation suggests that black holes emit thermal radiation, seemingly erasing the information contained in the matter that fell into the black hole~\cite{Hawking1975, PhysRevD.14.2460}.
	This conclusion is in tension with the unitarity of quantum mechanics.
	
	A key development in this direction is the Page curve~\cite{PhysRevLett.71.1291, PhysRevLett.71.3743, Penington2022}.
	For unitary black-hole evaporation, the entanglement entropy of the emitted radiation initially increases and then decreases after the Page time.
	Beyond the Page time, information entering the black hole can in principle be recovered from the radiation.
	This observation turns the information-loss problem into a more operational question: (1) how quickly, and (2) by what procedure, can newly infalling information be recovered from Hawking radiation?
	
	The Hayden--Preskill (HP) thought experiment addresses the first question (1)~\cite{Hayden2007}.
	It considers an old black hole that is already highly entangled with its early radiation after the Page time.
	When Alice throws a quantum diary into the black hole, the diary information is rapidly scrambled into the black-hole degrees of freedom.
	The thought experiment states that Bob can recover the diary after the scrambling time if he has access to the early radiation and collects a few additional qubits of late radiation.
	In this sense, an old black hole acts as a quantum information mirror.
	
	The Yoshida–Kitaev (YK) protocol addresses the second question (2) by providing an explicit decoder for the HP thought experiment~\cite{yoshida2017efficientdecodinghaydenpreskillprotocol,PhysRevX.9.011006}.
	Bob prepares an auxiliary copy of the input system and applies the conjugated scrambling dynamics jointly to the auxiliary system and early radiation.
	In the probabilistic version, the decoder postselects on an Einstein--Podolsky--Rosen (EPR) projection between the late radiation and its counterpart.
	Conditioned on successful postselection, the diary information is teleported to an auxiliary reference system.
	The YK protocol therefore turns the HP thought experiment into an operational recovery procedure and has, together with its variants, been implemented in several quantum-device experiments~\cite{Landsman2019,PhysRevX.11.021010,Kim2023,PhysRevD.109.044005,PhysRevD.110.026010,PhysRevResearch.7.023032,tm83-sxpm}.
	
	The YK protocol is quantitatively characterized by two quantities: the postselection probability ($P$) and the conditional fidelity ($F$).
	The probability $P$ measures how often the EPR projection succeeds, whereas $F$ measures how faithfully the successful branch recovers the diary information.
	Successful recovery requires both a non-negligible $P$ and a large $F$.
	These quantities are closely related to scrambling diagnostics, particularly out-of-time-order correlators (OTOCs)~\cite{Hosur2016,yoshida2017efficientdecodinghaydenpreskillprotocol,PhysRevX.9.011006}.
	Thus, measuring both $P$ and $F$ provides a direct quantum-device test of scrambling-induced information recovery~\cite{Landsman2019}.
	
	
	Meanwhile, many theoretical analyses of the HP and YK protocols model the scrambling dynamics as a Haar-random unitary.
	While this approach simplifies the analysis of recovery, it does not describe the real-time dynamics through which information is scrambled and subsequently recovered.
	Since the recoverability of the protocols is quantified by OTOCs, which are intrinsically time-dependent, realizing the protocols via Hamiltonian time evolution gives direct access to their dynamical properties~\cite{PhysRevB.98.014309,PhysRevResearch.2.043024,PhysRevD.104.074518,PhysRevResearch.6.L022021,Mao2026}.
	This Hamiltonian formulation is also well suited to quantum simulation, where the real-time evolution can be implemented directly.
	
	In addition, the old black hole is often assumed to be maximally entangled with the early radiation subsystem, forming an EPR state.
	For a given system, this corresponds to the infinite-temperature limit, which simplifies the analysis but does not capture the thermal nature of black-hole dynamics.
	A more realistic setup replaces the EPR state with a thermofield-double (TFD) state at finite temperature~\cite{PhysRevD.103.046004, PhysRevResearch.6.L022021,Mao2026}, making both $P$ and $F$ temperature dependent~\cite{PhysRevResearch.2.043024} and recovery generally more difficult than in the infinite-temperature limit~\cite{PhysRevD.106.046011}.
	
	While several studies have investigated HP recovery using Hamiltonian dynamics or at finite temperature separately, a few works have considered both setups together~\cite{PhysRevResearch.2.043024,10.1093/ptep/ptad147,PhysRevResearch.6.L022021}.
	To model black-hole information recovery in a more realistic setting, it is important to consider both Hamiltonian dynamics and finite temperature.
	Therefore, in this paper, we analyze finite-temperature HP recovery generated by Hamiltonian dynamics.
	
	However, such a Hamiltonian realization at finite temperature involves a subtle ambiguity.
	In conventional constructions, the diary is directly appended to the black hole before scrambling~\cite{PhysRevResearch.2.043024, PhysRevResearch.6.L022021}.
	The subsequent dynamics therefore acts on an enlarged Hilbert space and requires specifying an extended scrambling Hamiltonian beyond the one defining the initial TFD state.
	This construction alone does not uniquely prescribe how this extended Hamiltonian is related to the original black-hole Hamiltonian or how the diary is coupled to the black hole.
	
	To avoid this ambiguity, we introduce a SWAP gate into the protocol; instead of appending the diary to the old black hole, we exchange it with a chosen qubit in the fixed black-hole Hilbert space.
	The same Hamiltonian can then be used both to define the TFD state and to generate the subsequent scrambling.
	
	In choosing the model Hamiltonian, we require a many-body system with sufficiently strong scrambling for information recovery.
	Here, we consider the Sachdev--Ye--Kitaev (SYK) model, a strongly interacting system of $N$ Majorana fermions with random all-to-all interactions~\cite{PhysRevLett.70.3339,Kitaev2015}.
	The SYK model exhibits quantum chaos and rapid operator growth and is related to nearly AdS$_2$ black holes in the large-$N$ and low-energy limit~\cite{Maldacena2016,PhysRevD.94.106002,PhysRevLett.126.030602,10.1093/ptep/ptw124,PhysRevLett.117.111601}.
	It has therefore been widely used to study scrambling and black-hole-inspired quantum protocols~\cite{Gao2017,Gao2021,PhysRevX.12.031013,Jafferis2022,PRXQuantum.4.010320,PRXQuantum.4.010321,byun2026quantumsimulationtraversablewormholeinspiredquantum}.
	Sparse SYK models further reduce the circuit cost while retaining the relevant dynamics, making them suitable for quantum simulation~\cite{xu2020sparsemodelquantumholography,PhysRevD.103.106002,Caceres2021,PhysRevB.107.L081103,Granet2026}.
	These features make the SYK models natural settings for Hamiltonian HP recovery~\cite{10.1093/ptep/ptad147, PhysRevResearch.6.L022021}, and we therefore adopt them as toy models for the present protocol.

	\begin{table*}[t]
		\centering
		\setlength{\tabcolsep}{10pt}
		\renewcommand{\arraystretch}{1.3}
		\begin{tabular}{c|ccc}
			& Infinite-$T$ appended YK & Finite-$T$ appended YK & Finite-$T$ SWAP-injected YK \\
			\hline
			Diary injection & append $A$ to $B$ & append $A$ to $B$ & SWAP $A\leftrightarrow b\in B$ \\
			Hilbert space & enlarged, $d_{A}d_{B}=d_{C}d_{D}$ & enlarged, $d_{A}d_{B}=d_{C}d_{D}$ & fixed, $d_{B}=d_{C}d_{D}$ \\
			Scrambling dynamics & Haar or $H_{AB}$ & Haar or extended $H_{AB}$ & single fixed $H_{B}$ \\
			Collected radiation & $B'D$ & $B'D$ & $B'D$ and post-SWAP $A$ \\
			Thermal factors in $P$, $F$ & none & $\eta_{\beta}$ & $\eta_{\beta}$ and $\xi_{\beta}$ \\
		\end{tabular}
		\caption{\label{table1}
			Comparison of the conventional, appended Yoshida--Kitaev (YK) decoder and the SWAP-injected protocol used in this work.
			The appended decoder injects the diary $A$ by enlarging the Hilbert space of scrambling operations, so its finite-temperature version requires specifying an extended scrambling Hamiltonian $H_{AB}$ beyond the one preparing the TFD state.
			The SWAP injection instead exchanges $A$ with a chosen qubit $b$ in the initial black-hole system $B$, so that a single fixed Hamiltonian $H_{B}$ both defines the TFD state and generates the scrambling on the fixed Hilbert space.
			This forces the expelled qubit to be collected as radiation $A$ together with the late radiation $D$, producing the injection-dependent thermal factor $\xi_\beta$ absent in the conventional decoder ($\eta_{\beta}$ and $\xi_{\beta}$ will be described in Sec.~\ref{subsec:IIID}).
		}
	\end{table*}

	The main contributions of this work are threefold.
	First, the SWAP injection provides \textit{a single-Hamiltonian realization of finite-temperature HP recovery on a fixed Hilbert space} without requiring an enlarged post-injection Hamiltonian.
	Second, including the expelled black-hole qubit as initial radiation introduces an additional thermal factor $\xi_{\beta}$, absent from the appended decoder, that modifies both $P$ and $F$.
	Third, the SWAP construction naturally supplies a reference circuit for protocol-adapted error mitigation of both observables in a quantum-hardware implementation.
	
	We substantiate these contributions by analyzing the SWAP-injected protocol with SYK models and then implementing a binary sparse version on quantum hardware.
	We compute $P$ and $F$, derive their exact early-time values, and estimate their late-time saturation values under a uniform-spreading assumption.
	In particular, the late-time values are controlled by two thermal factors determined by $\rho_{\beta}^{1/2}$ and its reduced operators.
	The structural differences from the appended decoder are summarized in Table~\ref{table1}.
	
	For quantum-hardware implementation, we then adopt the binary sparse SYK model and implement the protocol on an IBM superconducting quantum computer with $N=8$ at two temperatures.
	Although the circuit depth suppresses both $P$ and $F$, the qualitative recovery behavior remains clear.
	The protocol-adapted reference circuit substantially restores the postselection probability and partially restores the conditional fidelity.
	
	The rest of this paper is organized as follows.
	In Sec.~\ref{sec:finite_temperature_protocol}, we introduce the SWAP-injected HP thought experiment and the probabilistic YK decoder at finite temperature.
	In Sec.~\ref{sec:syk_realization}, we realize the protocol using SYK models, derive the saturation behavior, and compare the recovery dynamics with a scrambling diagnostic.
	In Sec.~\ref{sec:quantum_computer_simulation}, we demonstrate the quantum-computer implementation of the finite-temperature decoder and analyze hardware noise using a protocol-adapted error-mitigation scheme.
	We conclude in Sec.~\ref{sec:discussion}.
	The analytical derivations are provided in the appendices.

	\section{\label{sec:finite_temperature_protocol}Finite-temperature Hayden--Preskill recovery}
	We consider a finite-temperature version of the HP thought experiment and its explicit recovery protocol, the YK protocol.
	Throughout this paper, we denote the number of qubits in a system $X$ by $N_{X}$ and the corresponding Hilbert-space dimension by $d_{X}=2^{N_{X}}$.
	
	\subsection{\label{subsec:hp_protocol}SWAP-injected Hayden--Preskill thought experiment}
	We first briefly review the conventional HP protocol.
	We consider an initial system consisting of the diary $A$, prepared by Alice, its reference $R$, the initial black-hole system $B$, and the early radiation $B'$.
	We take $N_{A} = N_{R}$ and $N_{B} = N_{B'}$.
	Then, the systems $A$ and $R$ are initially maximally entangled as
	\begin{equation}
		\ket{\mathrm{EPR}}_{RA} = \frac{1}{\sqrt{d_{A}}}\sum_{n=0}^{d_{A} - 1}\ket{n}_{R}\ket{n}_{A},
	\end{equation}
	and the initial black hole $B$ is also maximally entangled with the early radiation $B'$.
	This corresponds to the infinite-temperature limit of the entanglement between $B$ and $B'$.
	To incorporate finite temperature $T$, we replace the EPR state on $BB'$ by the TFD state
	\begin{equation}
		\label{eq:TFD}
		\ket{\mathrm{TFD}}_{BB'}
		=
		\frac{1}{\sqrt{Z_{\beta}}}
		\sum_{n}
		e^{-\beta E_{n}/2}
		\ket{n}_{B}\ket{n^{*}}_{B'},
	\end{equation}
	where $\beta=1/T$ and $Z_{\beta}=\sum_{n}e^{-\beta E_{n}}$ for a given Hamiltonian $H_{B}$ on $B$, with the corresponding Hamiltonian on $B'$ taken to be $H_{B'} = H_{B}$.
	The full initial state is therefore
	\begin{equation}
		\label{eq3}
		\ket{\Psi_{0}} = \ket{\mathrm{EPR}}_{RA}\otimes\ket{\mathrm{TFD}}_{BB'}.
	\end{equation}
	We assume that this initial state is prepared at $t = 0$.
	Throughout this paper, we set $N_{A} = N_{R} = 1$ for simplicity, so that the EPR state reduces to the Bell state $\ket{\Phi}_{RA} = \frac{1}{\sqrt{2}}(\ket{00} + \ket{11})_{RA}$. 
	The generalization to $N_{A} > 1$ is straightforward.

	We assume that the black-hole dynamics is simply described by a scrambling unitary operation $U$.
	In the conventional HP thought experiment, $A$ is directly appended to $B$, and the enlarged system $AB$ is evolved by the unitary operation $U_{AB}$, which is modeled as a Haar-random unitary.
	The output system is then partitioned into the remaining black hole $C$ and the late radiation $D$, satisfying $d_{A}d_{B} = d_{C}d_{D}$.
	Assuming that the outside observer Bob has access to the full radiation $B'D$, the recovery problem is to determine whether the information initially contained in $A$ can be reconstructed from the collected radiation.
	
	The implication of the thought experiment is that, if the dynamics $U_{AB}$ is sufficiently scrambling, then the diary information becomes recoverable from the radiation $B'D$ once $R$ is decoupled from $C$.
	Quantitatively, at $\beta\to0$, this is estimated by~\cite{doi:10.1142/S1230161208000043, Hayden2007,10.1093/ptep/ptad147}
	\begin{equation}
		\label{eq:L1norm}
		\mathbb{E}_{U}\left(\left\|\rho_{RC}-\rho_{R}\otimes\rho_{C}\right\|_{1}^{2}\right) \leq \left(\dfrac{d_{A}}{d_{D}}\right)^{2},
	\end{equation}
	where the average is taken over Haar-random unitaries and the $L_{1}$ norm is defined by $\|A\|_{1}=\Tr\sqrt{A^{\dagger}A}$.
	This implies that Bob needs to collect only slightly more late radiation qubits than the number of diary qubits after the scrambling time.
	Thus the information is not lost, but is rapidly transferred to the radiation degrees of freedom.
	At finite $T$, the modification of \eqref{eq:L1norm} is suggested in Ref.~\cite{PhysRevD.106.046011}.

	\begin{figure*}[tb]
		\includegraphics[width=0.6\linewidth]{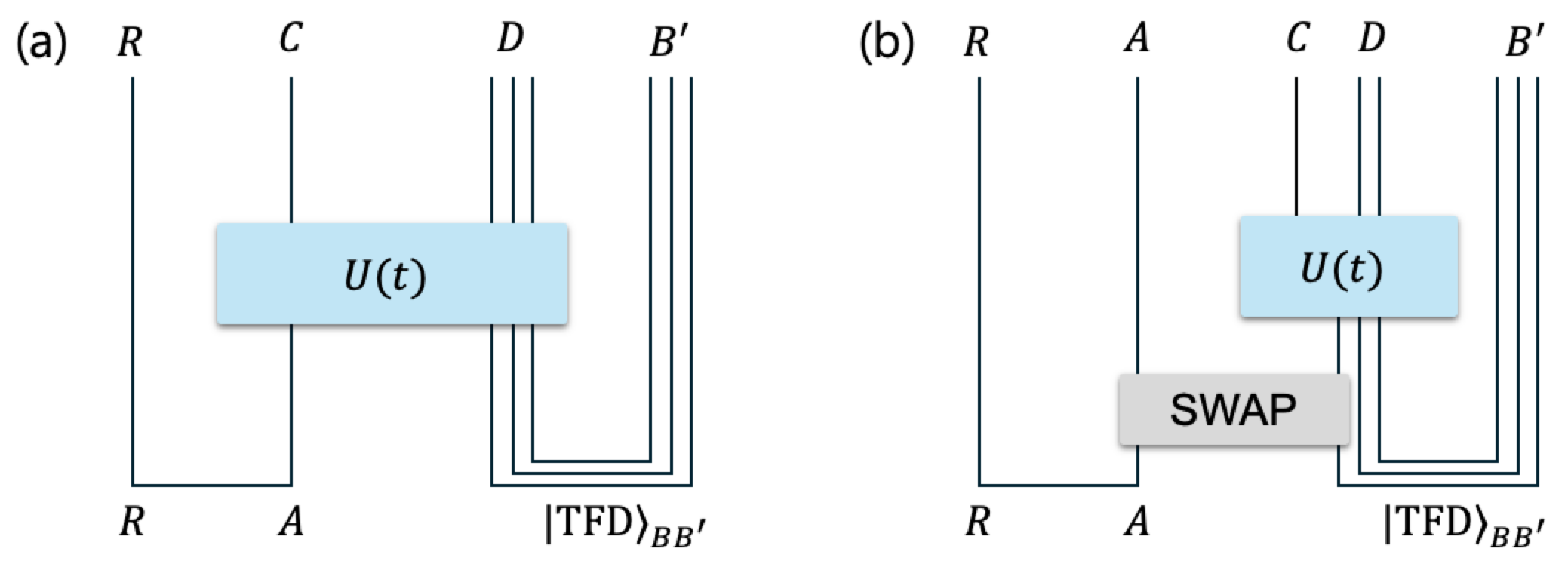}
		\caption{
			Quantum circuits for the finite-temperature Hayden--Preskill thought experiment.
			(a) Conventional representation, where the diary $A$ is appended to the old black hole $B$ and the enlarged system $AB$ is scrambled by $U(t)$.
			(b) SWAP-injected protocol, where the diary is injected into a chosen qubit $b\in B$ via a SWAP gate, so that the TFD state and the subsequent scrambling dynamics are defined by the same Hamiltonian on the fixed Hilbert space $\mathcal{H}_{B}$.
		}
		\label{fig:circuit}
	\end{figure*}

	For a Hamiltonian realization of the protocol, we consider the case in which $U_{AB}$ is the time-evolution operator $U_{AB}(t)$.
	That is, we assume that the black-hole dynamics is described by $H_{AB}$ acting on $\mathcal{H}_{A}\otimes\mathcal{H}_{B}$, such that $U_{AB}(t) = e^{-iH_{AB}t}$.
	The corresponding HP protocol is shown in Fig.~\ref{fig:circuit}(a).
	
	At finite temperature, however, the initial TFD state in \eqref{eq:TFD} is defined by the Hamiltonian $H_{B}$, whereas the subsequent evolution is generated by $H_{AB}$ on the enlarged Hilbert space~\cite{PhysRevResearch.2.043024, PhysRevResearch.6.L022021}.
	The conventional construction does not specify how $H_{AB}$ should extend $H_{B}$ or how the diary should be coupled to the original black hole.\footnote{Although the diary can instead be encoded through an isometric insertion~\cite{10.1093/ptep/ptad147}, such an insertion is not unitary on a fixed Hilbert space.}
	
	To avoid this ambiguity, we introduce the modified HP protocol shown in Fig.~\ref{fig:circuit}(b).
	Instead of enlarging $\mathcal{H}_{B}$, we exchange a qubit on $A$ with a chosen qubit $b\in B$ by a SWAP operation $S_{Ab}$, defined by
	\begin{equation}
		\label{eq6}
		S_{Ab}\,\ket{i}_{A}\ket{j}_{b} = \ket{j}_{A}\ket{i}_{b},
		\qquad
		i, j \in \{0, 1\}.
	\end{equation}
	The subsequent unitary scrambling is then generated by the same $H_{B}$ that defines the TFD state in \eqref{eq:TFD}, such that
	\begin{equation}
		\label{eq:post_infall_unitary}
		U_{B}(t) = e^{-iH_{B}t},
		\qquad 
		H_{B}\ket{n}=E_{n}\ket{n}.
	\end{equation}
	After the scrambling, $B$ is partitioned into $CD$, such that $d_{B} = d_{C}d_{D}$, which is distinct from the conventional HP thought experiment.
	Thus, the SWAP-injected protocol guarantees that both the initial TFD state and the black-hole dynamics are defined by the same Hamiltonian.

	In this modified protocol, the SWAP replaces a pre-existing black-hole qubit degree of freedom by the diary qubit and expels that degree of freedom into the external register $A$.
	We interpret this expelled qubit as initial radiation emitted at the injection step, so Bob collects $A$ together with $DB'$ to recover the diary.
	Including $A$ in the collected radiation also makes the comparison with the conventional protocol well defined.
	Although the SWAP-injected construction obeys $d_{B}=d_{C}d_{D}$ rather than $d_{A}d_{B}=d_{C}d_{D}$, including $A$ in the collected radiation allows the dimensions of the remaining black hole and the total radiation to be matched between the two protocols for fixed $d_{A}$ and $d_{B}$.
	
	This initial radiation is distinct from both the early radiation $B'$ and the late radiation $D$: $B'$ has already been emitted before the diary injection, whereas $D$ is obtained after the post-injection scrambling evolution $U_B(t)$.
	The register $A$, by contrast, contains the pre-existing black-hole degree of freedom expelled at the moment of diary injection and remains outside the subsequent evolution.
	This use of a unitary interchange is reminiscent of qubit-transport models of unitary black-hole evaporation, in which a black-hole degree of freedom is transferred to an outgoing Hawking radiation qubit~\cite{PhysRevD.97.066023}.
	Here, however, the SWAP serves instead as a diary-injection operation on a fixed Hilbert space, with the expelled qubit included in the collected radiation.

	\subsection{Probabilistic Yoshida--Kitaev decoder}
	We now describe the recovery process, first reviewing the original YK protocol and then its SWAP-injected modification.
	
	To initialize the YK protocol, Bob prepares auxiliary systems $A'$ and $R'$ with $d_{A'}=d_{A}$ and $d_{R'}=d_{R}$, which are copies of $A$ and $R$, respectively.
	Similar to the state on $RA$, the auxiliary pair is also initialized in an EPR state.
	Together with the state \eqref{eq3}, the full initial state is
	\begin{equation}
		\label{eq:initial_state}
		\ket{\mathrm{in}}=\ket{\mathrm{EPR}}_{RA}\otimes\ket{\mathrm{TFD}}_{BB'}\otimes\ket{\mathrm{EPR}}_{R'A'}.
	\end{equation}
	Assuming that Bob knows the complete black-hole dynamics, the conjugated unitary $U_{A'B'}^{*}(t)$ is applied on the decoder side $A'B'$.
	This is the conjugate of $U_{AB}(t)$ and maps $A'B'$ to the corresponding output systems $C'D'$ with $d_{C'} = d_{C}$ and $d_{D'} = d_{D}$.

	\begin{figure*}[tb]
		\includegraphics[width=0.85\linewidth]{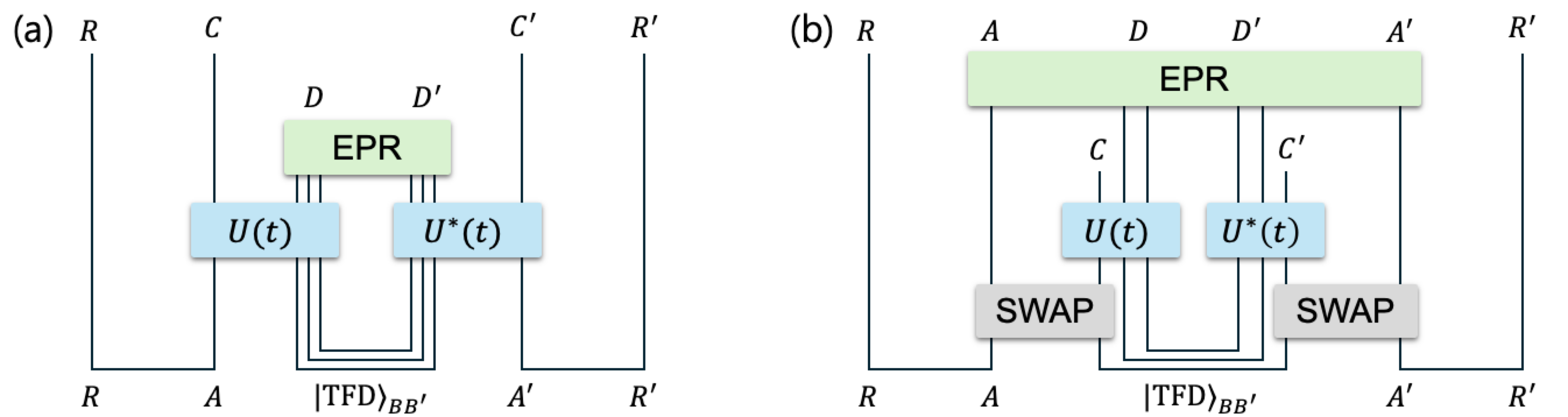}
		\caption{
			Quantum circuits for the finite-temperature Yoshida--Kitaev (YK) decoding protocol.
			(a) Conventional YK decoder, where $A'$ is appended to $B'$ and evolved by the conjugated unitary $U_{A'B'}^{*}(t)$ associated with the original scrambling unitary $U_{AB}(t)$.
			The EPR projection on $DD'$ implements the probabilistic decoder.
			(b) SWAP-injected protocol.
			The diary $A$ and the auxiliary diary $A'$ are injected into $B$ and $B'$ by the SWAP gates $S_{Ab}$ and $S_{A'b'}$, respectively.
			The two copies evolve under $U_{B}(t)$ and $U_{B'}^{*}(t)$, and the EPR projections on $AA'$ and $DD'$ define the postselection probability.
			Successful recovery is diagnosed by the conditional EPR fidelity between $R$ and $R'$.
		}
		\label{fig:circuit_2}
	\end{figure*}

	The YK protocol allows two distinct methods for information recovery: a probabilistic decoder and a deterministic decoder~\cite{yoshida2017efficientdecodinghaydenpreskillprotocol}.
	The probabilistic decoder postselects on a successful EPR projection between $D$ and $D'$.
	This measurement selects the desired decoding branch with a certain probability, whose late-time value is often of order $1/d_{A}^{2}$.
	If the projection succeeds and the resulting state has large conditional EPR fidelity between $R$ and $R'$, the diary information is interpreted as having been recovered; if the projection fails, the conditional recovery fidelity is not evaluated for that branch.
	This conventional YK probabilistic decoder is illustrated in Fig.~\ref{fig:circuit_2}(a).
	The deterministic decoder instead uses an explicit recovery operation, motivated by Grover's search algorithm~\cite{10.1145/237814.237866}.
	In this work, we focus on the probabilistic decoder.
	We probe the temperature-dependent postselection probability and the conditional decoding fidelity, denoted by $P_{\beta}(t)$ and $F_{\beta}(t)$.
	
	We now apply this recovery scheme to the SWAP-injected setup shown in Fig.~\ref{fig:circuit_2}(b).
	As described in Sec.~\ref{subsec:hp_protocol}, $A$ is injected into $B$ by exchanging it with the chosen qubit $b\in B$.
	On the decoder side, the auxiliary diary $A'$ is similarly exchanged with the corresponding qubit $b'\in B'$ via the SWAP gate $S_{A'b'}$.
	Afterward, Bob applies $U_{B'}^{*}(t)=e^{+iH_{B'}^{*}t}$ to $B'$, and the output is partitioned into $C'D'$.
	The pre-measurement state is then
	\begin{equation}
		\label{eq:premeasurement_state}
		|\widetilde{\mathrm{out}}(t)\rangle=U_{B}(t)U_{B'}^{*}(t)S_{Ab}S_{A'b'}\ket{\mathrm{in}}.
	\end{equation}
	Here, the qubits $b$ and $b'$ are taken to be symmetric in the YK decoder as verified in Sec.~\ref{subsec:IIIC}.
	
	Now Bob performs the EPR measurements required for postselection.
	Since $A$ after the SWAP is regarded as radiation, as described in Sec.~\ref{subsec:hp_protocol}, the EPR projections are performed on both $AA'$ and $DD'$.
	The postselection succeeds only when the EPR measurement returns the maximally entangled outcome $\ket{\mathrm{EPR}}_{AA'}\otimes\ket{\mathrm{EPR}}_{DD'}$.
	The corresponding EPR projector is
	\begin{equation}
		\label{eq:epr_projection}
		\Pi_{AA'}\Pi_{DD'} = \ket{\mathrm{EPR}}\bra{\mathrm{EPR}}_{AA'}\otimes\ket{\mathrm{EPR}}\bra{\mathrm{EPR}}_{DD'}.
	\end{equation}
	The postselection probability is then obtained from
	\begin{equation}
		\label{eq:postselection_probability}
		P_{\beta}(t)=\langle\widetilde{\mathrm{out}}(t)|\,\Pi_{AA'}\Pi_{DD'}\,|\widetilde{\mathrm{out}}(t)\rangle.
	\end{equation}
	Conditioned on successful EPR projection, the normalized postselected state is
	\begin{equation}
		\label{eq:postselected_state}
		\ket{\mathrm{out}(t)}=\frac{1}{\sqrt{P_{\beta}(t)}}\,\Pi_{AA'}\Pi_{DD'}\,|\widetilde{\mathrm{out}}(t)\rangle.
	\end{equation}
	Projecting on $AA'$ in addition to $DD'$ is the operational counterpart of collecting the initial radiation.
	Whereas the appended decoder postselects on $DD'$ alone, here the postselection also probes the thermal correlations carried by the expelled qubit now residing in $A$.
	As we will see in Sec.~\ref{subsec:IIID}, collecting this expelled qubit is the origin of the injected-site thermal factor ($\xi_{\beta}$ in \eqref{eq:eta_xi_main}) that distinguishes the SWAP-injected protocol from the conventional finite-temperature YK decoder~\cite{PhysRevD.106.046011} when the collected radiation is small.
	
	The recovery fidelity is evaluated by checking whether $R$ becomes maximally entangled with $R'$.
	Let $\rho_{RR'}(t)$ be the reduced density matrix of $\ket{\mathrm{out}(t)}$, obtained by tracing out all degrees of freedom except $R$ and $R'$.
	Then, the conditional fidelity is
	\begin{equation}
		\label{eq:recovery_fidelity}
		F_{\beta}(t)=\bra{\mathrm{EPR}}\rho_{RR'}(t)\ket{\mathrm{EPR}}_{RR'}.
	\end{equation}
	This directly measures whether the postselected branch recovers the diary information in $R'$.
	Therefore, successful recovery requires a non-negligible $P_{\beta}(t)$ together with a large $F_{\beta}(t)$.
	



	\section{\label{sec:syk_realization}SYK realization of finite-temperature recovery}
	In the YK decoder described above, we realize the scrambling dynamics using the SYK Hamiltonian.
	
	\subsection{\label{subsec:syk_model}Sachdev--Ye--Kitaev model}
	The $q$-body SYK Hamiltonian with $N$ strongly interacting Majorana fermions is defined by
	\begin{equation}
		\label{eq:syk_hamiltonian}
		H_{\rm SYK}
		=
		i^{q/2}
		\sum_{1\leq i_{1}<\cdots<i_{q}\leq N}
		J_{i_{1}\cdots i_{q}}
		\chi_{i_{1}}\cdots\chi_{i_{q}},
	\end{equation}
	where $\{\chi_{i},\chi_{j}\}=\delta_{ij}$.
	The random couplings are Gaussian-distributed with zero mean and variance
	\begin{equation}
		\label{eq:syk_variance_full}
		\left\langle J_{i_{1}\cdots i_{q}}\right\rangle = 0,
		\qquad
		\left\langle J_{i_{1}\cdots i_{q}}^{2}\right\rangle
		=
		\dfrac{J^{2}(q-1)!}{N^{q-1}}.
	\end{equation}
	In this paper, we set $q=4$ and $J=\sqrt{2}$.
	
	For numerical and hardware implementations, we represent the Majorana fermions by the Jordan--Wigner transformation
	\begin{equation}
		\chi_{2m-1}
		=
		\dfrac{Z^{\otimes(m-1)}X_{m}}{\sqrt{2}},
		\qquad
		\chi_{2m}
		=
		\dfrac{Z^{\otimes(m-1)}Y_{m}}{\sqrt{2}},
	\end{equation}
	with $m=1,\ldots,N/2$.
	Thus, the SYK model with $N$ Majorana fermions is implemented on a $2^{N/2}$-dimensional Hilbert space.
	We use the same $H_{\text{SYK}}$ with the same disorder realization to define the TFD state and the scrambling dynamics.
	
	The symmetry of the dynamics affects the recoverability of the HP protocol~\cite{PhysRevResearch.2.043164, PhysRevD.100.086001, PhysRevResearch.6.L022021, Nakata2023blackholesasclouded, tajima2022universallimitationquantuminformation}.
	The SYK$_{4}$ model has a parity symmetry, which can cause a deviation in the YK decoder because the dynamics cannot be fully ergodic in the full Hilbert space.
	For this reason, we choose the even-parity sector of the SYK Hamiltonian.
	That is, if the system size of the SYK model is $N$, we represent the systems $B$ and $B'$ as $N_B = N_{B'} = N/2 - 1$ qubit systems in the computational basis.
	Accordingly, all subsequent states, traces, and operations are defined within these even-parity qubit Hilbert spaces rather than in the full $2^{N/2}$-dimensional Hilbert space.
	In particular, $b \in B$ and $b' \in B'$ in \eqref{eq:premeasurement_state} denote qubits in this parity-reduced representation, so that the SWAP gates preserve the parity of $B$ and $B'$.
	
	To suppress finite-$N$ fluctuations, and because the SYK model involves random disorder, we consider disorder-averaged dynamical quantities, such as $\overline{P_{\beta}(t)}$ and $\overline{F_{\beta}(t)}$.
	Unless otherwise stated, these quantities are averaged over 20 disorder realizations.

	\subsection{\label{sec:IIIB}Decoupling and temperature dependence}
	For the SYK Hamiltonian, we consider the disorder-averaged squared $L_{1}$ norm on the left-hand side of \eqref{eq:L1norm}, which diagnoses whether $R$ is decoupled from $C$.
	
	Strictly, the Haar-averaged bound on the right-hand side of \eqref{eq:L1norm} is derived for the conventional HP protocol at $\beta\to0$.
	In the SWAP-injected construction, the relevant relation is $d_{B} = d_{C}d_{D}$.
	Also, our setup considers finite temperature so that the bound $(d_{A}/d_{D})^{2}$ on the right-hand side need not carry over. 
	We therefore use the left-hand side of \eqref{eq:L1norm} only as a qualitative decoupling diagnostic: its decay signifies that the diary initially on $A$ and injected into $b\in B$ is no longer stored in $C$, but has been transferred to the radiation system accessible to the decoder.
	This also implies that $R$ and $C$ are nearly uncorrelated.
	This qualitative recovery condition is consistent with the enhancement of the conditional fidelity after scrambling as we will show later.
	
	We take $N_{A}=1$, $N_{B}=7$, $N_{C}=5$, and $N_{D}=2$ for the $N=16$ SYK Hamiltonian.
	We evaluate the norm at $T=0.01, 0.1, 0.2, 0.5, 1$, and $10$.
	Unless otherwise stated, we use the same temperatures for the numerical evaluation of the other quantities presented below.

	\begin{figure}[tb]
		\centering
		\includegraphics[width=0.8\linewidth]{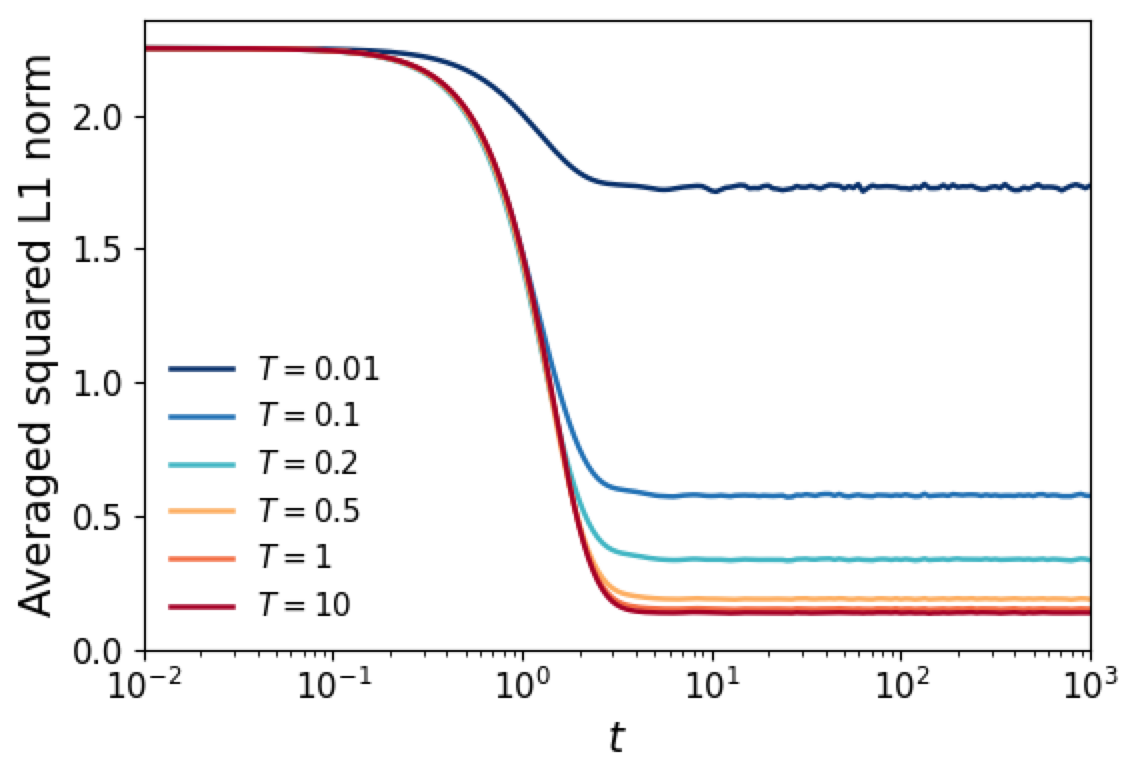}
		\caption{
			\label{fig:L1norm}
			Decoupling diagnostic for the finite-temperature Hayden--Preskill (HP) protocol with $N_{A} = 1$, $N_{B} = 7$, $N_{C} = 5$, and $N_{D} = 2$.
			The time evolution is generated by the $N = 16$ SYK Hamiltonians.
			The disorder-averaged squared $L_{1}$ norm is shown for $T = 0.01, 0.1, 0.2, 0.5, 1$, and $10$ over 20 disorder realizations.
			The decrease of this quantity at higher temperature indicates stronger decoupling between the reference $R$ and the remaining black hole $C$, which is consistent with the original HP thought experiment.
		}
	\end{figure}

	As shown in Fig.~\ref{fig:L1norm}, the decoupling norm decreases within a finite time window, and the late-time saturation decreases as $T$ increases.
	When $\beta\to0$, the thermal state $\rho_{\beta} = e^{-\beta H}/Z_{\beta}$ approaches the maximally mixed state $I_{B}/d_{B}$.
	Thus, the protocol approaches the ordinary HP protocol, where the standard decoupling argument applies directly; that is, the squared $L_{1}$ norm decays nearly to zero after scrambling.
	By contrast, the increase of the squared $L_{1}$ norm at lower $T$ reflects the fact that the TFD state has dominant support in the low-energy sector of the Hilbert space, so the injected diary information is less efficiently decoupled from the remaining black-hole subsystem.

	\subsection{\label{subsec:IIIC}Postselection probability and recovery fidelity}
	We numerically evaluate the postselection probability and fidelity for $N = 16$ SYK Hamiltonians for the SWAP-injected YK protocol.
	Throughout the numerical analysis, we choose the same injection site in the two copies, taking $b$ and $b'$ to be the first qubits of $B$ and $B'$, respectively.
	Under the output partitions $B=CD$ and $B'=C'D'$, these qubits belong to $C$ and $C'$, respectively.

	\begin{figure*}[tb]
		\centering
		\includegraphics[width=0.75\linewidth]{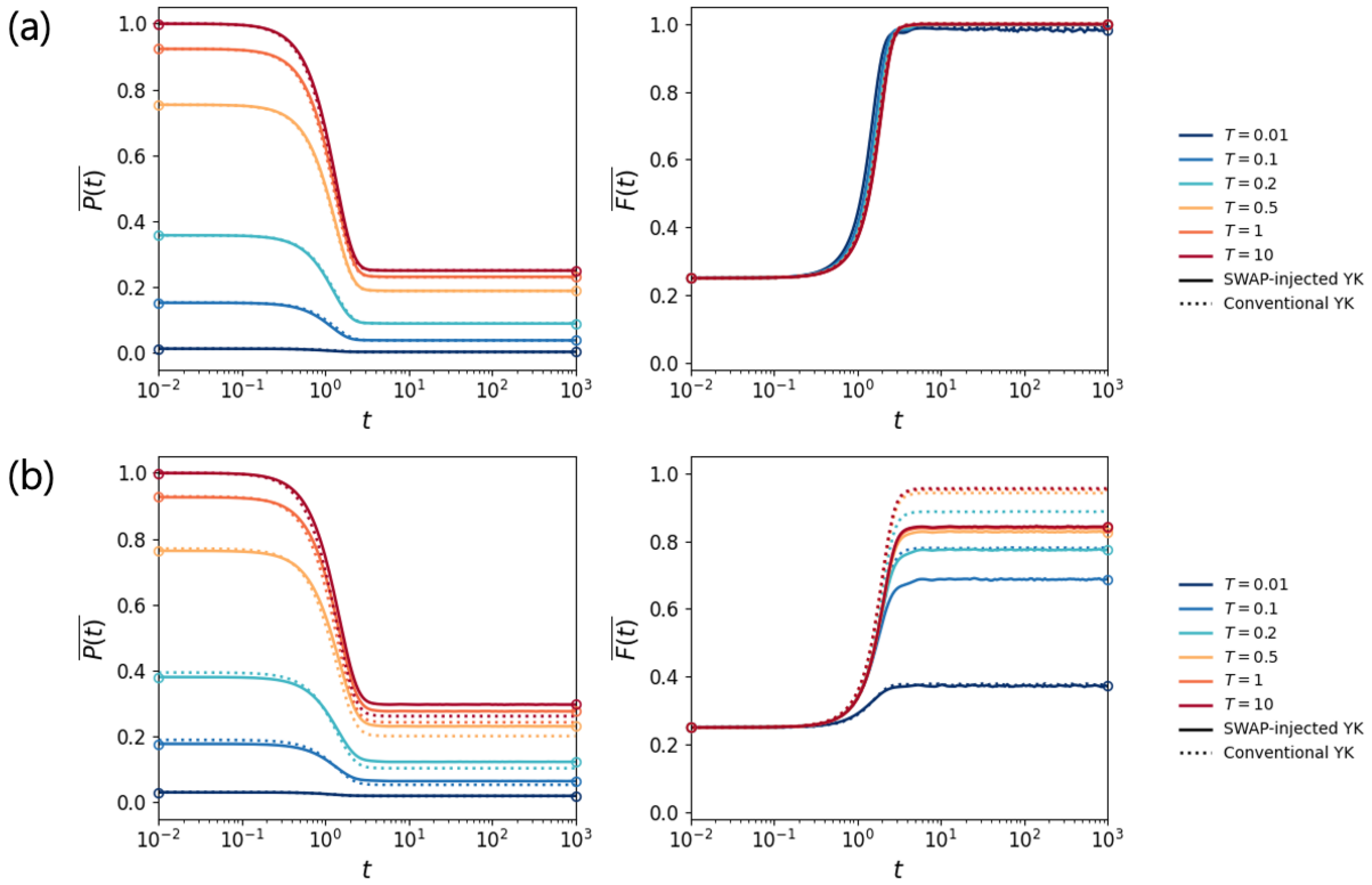}
		\caption{
			\label{fig:T_dependence}
			Finite-temperature Yoshida--Kitaev (YK) decoding dynamics with an $N=16$ SYK Hamiltonian for (a) $N_{A}=1$, $N_{B}=7$, $N_{C}=1$, and $N_{D}=6$, and (b) $N_{A}=1$, $N_{B}=7$, $N_{C}=5$, and $N_{D}=2$.
			The postselection probability $\overline{P_{\beta}(t)}$ and the conditional fidelity $\overline{F_{\beta}(t)}$ are shown for $T=0.01,0.1,0.2,0.5,1,$ and $10$, averaged over 20 disorder realizations.
			We show the SWAP-injected protocol by solid curves and the conventional YK protocol by dashed curves at matched output dimensions.
			For the latter, we take (a) $N_D=7$ and (b) $N_D=3$, while keeping the other system sizes fixed.
			The markers at the left and right edges denote the analytical early-time values and late-time saturation estimates for the SWAP-injected YK protocol.
		}
	\end{figure*}

	We first consider the case in which $N_{C}=1$ and $N_{D}=6$, such that $N_{D}$ takes its largest allowed value.
	As shown in Fig.~\ref{fig:T_dependence}(a), $\overline{P_{\beta}(t)}$ decreases during the time evolution, while $\overline{F_{\beta}(t)}$ increases over the same time window, consistently with the decoupling diagnostic discussed in Sec.~\ref{sec:IIIB}.
	At high $T$, their late-time saturation values approach $1/d_{A}^{2}$ and unity, respectively, as in the conventional YK protocol.
	As $T$ decreases, $\overline{P_{\beta}(t)}$ is further suppressed, whereas $\overline{F_{\beta}(t)}$ remains relatively insensitive to $T$.
	Thus, when the radiation subsystem is sufficiently large, temperature mainly affects the postselection probability rather than the conditional fidelity.
	Physically, Bob can recover the diary information with nearly unit fidelity even at low temperature, although the successful postselection becomes less probable.
	Figure~\ref{fig:T_dependence}(a) also shows that the SWAP-injected protocol (solid) exhibits probability and fidelity similar to those of the conventional finite-temperature YK protocol (dashed).\footnote{For the appended decoder, the TFD preparation and scrambling dynamics are averaged over independent disorder realizations.}
	
	To examine the dependence on the radiation subsystem size, we next consider the case $N_{C}=5$ and $N_{D}=2$, for which the decoder has access to fewer late-radiation degrees of freedom.
	The results are shown in Fig.~\ref{fig:T_dependence}(b).
	In this case, $\overline{F_{\beta}(t)}$ becomes more sensitive to $T$, and its high-temperature saturation value remains below unity.
	By contrast, $\overline{P_{\beta}(t)}$ remains similar to that obtained for $N_{D}=6$.
	These results show that reducing the radiation subsystem primarily suppresses the conditional fidelity, while having a much weaker effect on the postselection probability.
	
	These two regimes also reveal where the SWAP-injected construction departs from the conventional appended YK decoder.
	For a large radiation subsystem, shown in Fig.~\ref{fig:T_dependence}(a), the solid and dashed curves nearly coincide, so the SWAP injection reproduces the conventional finite-temperature YK behavior. 
	On the other hand, for a small radiation subsystem, shown in Fig.~\ref{fig:T_dependence}(b), the two protocols visibly separate at high temperature: the conditional fidelity of the SWAP-injected protocol no longer saturates to unity at high $T$.
	In Sec.~\ref{subsec:IIID}, we will provide an explicit analysis of this distinction.

	\begin{figure*}[tb]
		\centering
		\includegraphics[width=0.75\linewidth]{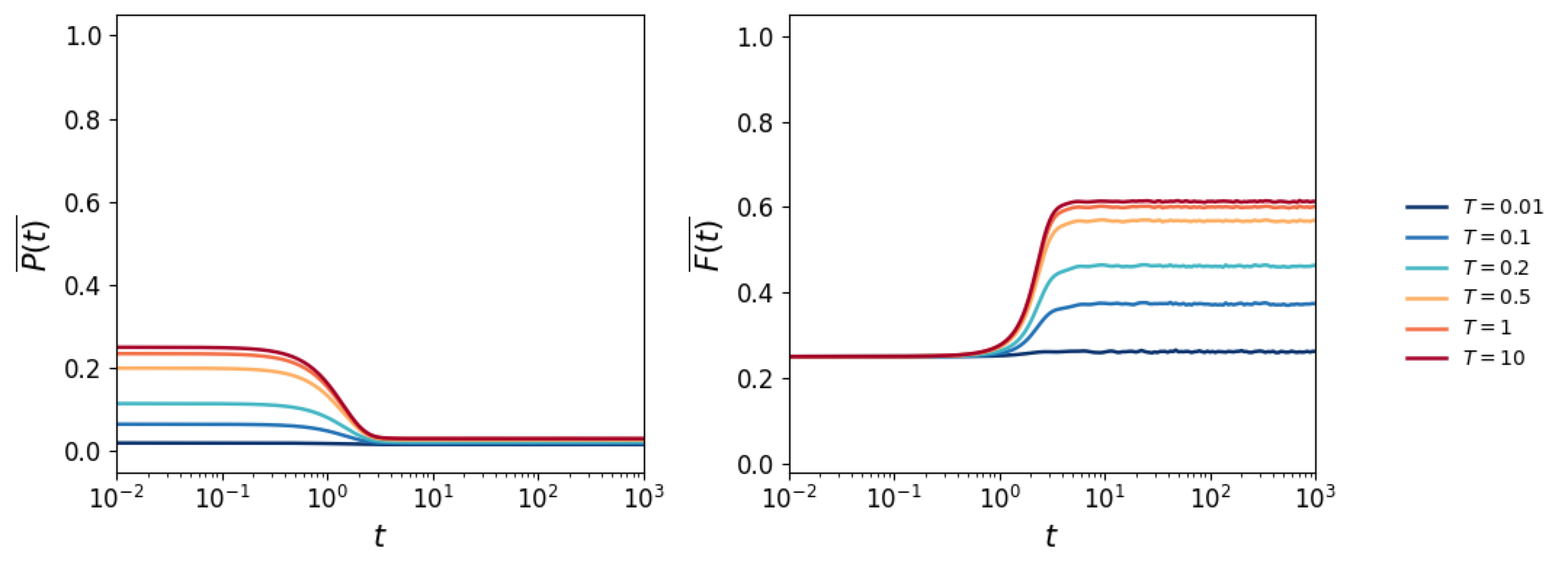}
		\caption{
			\label{fig:out_of_injection}
			Effect of mismatched injection positions in the SWAP-injected Yoshida--Kitaev (YK) decoder.
			We set $N_{A}=1$, $N_{B}=7$, $N_{C}=5$, and $N_{D}=2$.
			The original diary $A$ and the auxiliary diary $A'$ are injected into different qubit positions in $B$ and $B'$, respectively.
			We choose $b = 0$ and $b' = 1$.
			This mismatch suppresses both postselection probability and conditional fidelity.
		}
	\end{figure*}

	We also show that the decoder-side injection qubit must be chosen as the mirror copy of the injection qubit in $B$.
	When the same qubit position in $B$ and $B'$ is used for $S_{Ab}$ and $S_{A'b'}$, the decoder exhibits the expected dynamics, with a large $F_{\beta}(t)$ at high $T$.
	By contrast, as shown in Fig.~\ref{fig:out_of_injection}, when $A'$ is injected into a different qubit position in $B'$, both $P_{\beta}(t)$ and $F_{\beta}(t)$ are strongly suppressed.
	For this reason, we use the same injection position for the two SWAP gates throughout the rest of this work.
	
	\subsection{\label{subsec:IIID}Saturation and scrambling dynamics}
	We derive exact early-time values and analytic expressions for the late-time saturation values of $\overline{P_{\beta}(t)}$ and $\overline{F_{\beta}(t)}$ for the SWAP-injected YK protocol.
	
	\vspace{5pt}
	\noindent
	\textbf{Late-time saturation.}
	We first focus on $\overline{P_{\beta}(t)}$, whose detailed derivation is given in Appendix~\ref{appendix:B}.
	
	At finite $T$, the estimate below is obtained by assuming \textit{uniform} spreading of the traceless part of the thermal operator in the scrambled regime.
	More explicitly, we decompose the operator relevant to the dynamics, denoted by $X(t)$, into its identity and traceless components as $X(t) = X_{I}(t) + X_{0}(t)$, with $\Tr X_{0}(t) = 0$.
	It turns out that the identity contribution gives a closed form.
	The traceless component is assumed to spread uniformly over the basis after scrambling and disorder averaging.
	That is, if $X_{0}(t)$ is defined on $B=CD$, this refers to
	\begin{equation}
		\label{eq18}
		\overline{\left\|\Tr_{D}\!\left[X_{0}(t)\right]\right\|_{F}^{2}}
		\approx
		\dfrac{d_{D}(d_C^2-1)}{d_B^2-1}
		\overline{\left\|X_{0}\right\|_{F}^{2}}
	\end{equation}
	at late $t$, where $X_{0}(t) = U(t)X_{0}U^{\dagger}(t)$ and $\|\cdots\|_F$ denotes the Frobenius norm $\|\mathcal{O}\|_F^2=\Tr(\mathcal{O}^{\dagger}\mathcal{O})$.
	Here the factor $d_C^2-1$ counts the traceless operator directions on $C$ that survive the partial trace over $D$, while $d_B^2-1$ is the total number of traceless operator directions on $B$.
	We provide a numerical coefficient-level diagnostic of this uniform-spreading assumption in Appendix~\ref{appendix:isotropic_diagnostic}.
	
\begin{figure}[tb]
		\centering
		\includegraphics[width=0.7\linewidth]{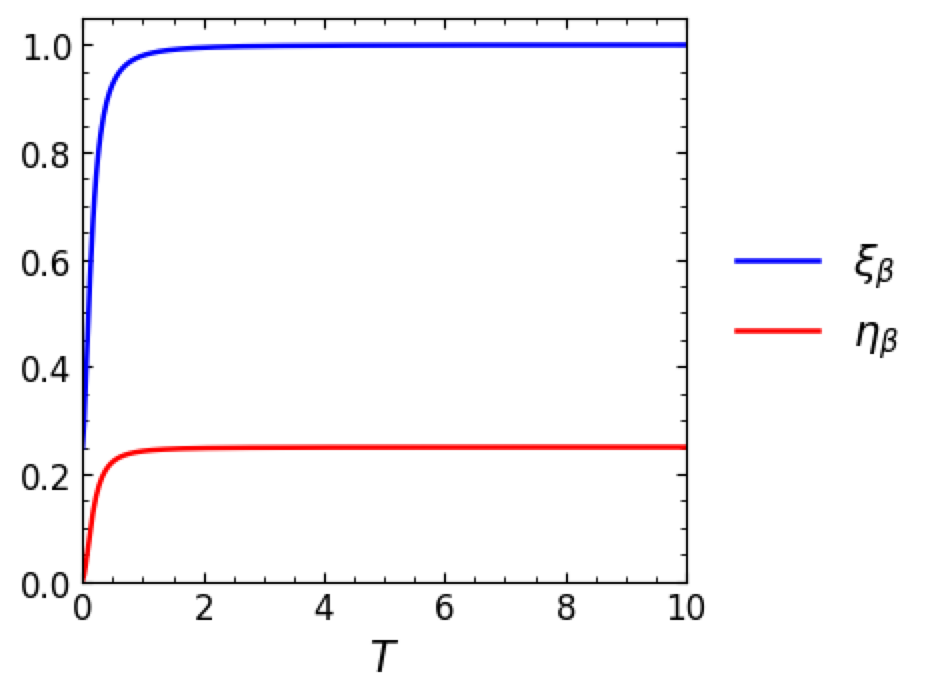}
		\caption{
			\label{fig:eta_beta}
			The thermal quantities $\eta_{\beta}$ and $\xi_{\beta}$ for $N_{A} = 1$, $N_{B} = 7$, $N_{C}=5$, and $N_{D}=2$.
			In the high-$T$ limit, they approach $\eta_{\beta\to0}=1/d_A^2$ and $\xi_{\beta\to0}=1$.
		}
	\end{figure}

	Following the derivation in Appendix~\ref{appendix:B}, the above uniform-spreading assumption yields the following analytic expression for the disorder-averaged finite-temperature saturation value:
	\begin{equation}
		\label{eq:P_sat_beta}
		\overline{P}_{\beta}
		\approx
		\eta_{\beta}+\kappa\left(\xi_{\beta}-\eta_{\beta}\right),
		\qquad
		\kappa := \dfrac{d_{C}^{2} - 1}{d_{B}^{2} - 1},
	\end{equation}
	where we have defined two thermal factors
	\begin{equation}
		\label{eq:eta_xi_main}
		\eta_{\beta}
		:=
		\dfrac{1}{d_A^2d_B}
		\overline{\left|\Tr\rho_{\beta}^{1/2}\right|^{2}},
		\qquad
		\xi_{\beta}
		:=
		\dfrac{1}{d_A}
		\overline{\left\|\Tr_{b}\rho_{\beta}^{1/2}\right\|_{F}^{2}}.
	\end{equation}
	Here, $b$ denotes the injected qubit inside $B$.
	The behavior of $\eta_{\beta}$ and $\xi_{\beta}$ is plotted in Fig.~\ref{fig:eta_beta}.
	We note that $\xi_{\beta}$ depends explicitly on the injected qubit through the partial trace over it, reflecting the distinctive structure of the SWAP-injected YK protocol.

	In the infinite-temperature limit, we have $\eta_{\beta}\to1/d_A^2$ and $\xi_{\beta}\to1$.
	Then, we find that the late-time saturation value is
	\begin{equation}
		\label{eq:P_sat}
		\overline{P}_{\beta\to0}\approx\dfrac{1}{d_A^2}+\kappa\left(1 - \dfrac{1}{d_{A}^{2}}\right).
	\end{equation}
	When $N_{D}$ is large enough, since $d_{B} = d_{C}d_{D}$, the second term scales as $O(d_{D}^{-2})$, which is consistent with the case of the original YK protocol~\cite{Hosur2016, PhysRevX.9.011006}.
	
	The same analysis also gives the late-time saturation of $\overline{F_{\beta}(t)}$.
	The detailed derivation is given in Appendix~\ref{appendix:C}.
	We consider the joint probability for successful EPR projection on $AA'$, $DD'$, and $RR'$, given by
	\begin{equation}
		Q_{\beta}(t)=\langle\widetilde{\mathrm{out}}(t)|\Pi_{AA'}\Pi_{DD'}\Pi_{RR'}|\widetilde{\mathrm{out}}(t)\rangle.
	\end{equation}
	Then the conditional fidelity is
	\begin{equation}
		\label{eq23}
		F_{\beta}(t) = \dfrac{Q_{\beta}(t)}{P_{\beta}(t)}.
	\end{equation}
	At late times, the disorder-averaged joint probability is estimated as
	\begin{equation}
		\label{eq:Q_sat_beta}
		\overline{Q}_{\beta}
		\approx
		\eta_{\beta}
		+
		\kappa\left(\dfrac{\xi_{\beta}}{d_A^2}-\eta_{\beta}\right).
	\end{equation}
	Therefore,
	\begin{equation}
		\label{eq:F_sat_beta}
		\overline{F}_{\beta}
		\approx
		\dfrac{
			\eta_{\beta}
			+
			\kappa\left(\dfrac{\xi_{\beta}}{d_A^2}-\eta_{\beta}\right)
		}{
			\eta_{\beta}
			+
			\kappa\left(\xi_{\beta}-\eta_{\beta}\right)
		} = 1 - \kappa\left(1 - \dfrac{1}{d_{A}^{2}}\right)\dfrac{\xi_{\beta}}{\overline{P}_{\beta}}.
	\end{equation}
	In the infinite-temperature limit, $\eta_{\beta}\to1/d_A^2$ and $\xi_{\beta}\to1$, so \eqref{eq:F_sat_beta} reduces to the usual YK relation~\cite{PhysRevX.9.011006}
	\begin{equation}
		\label{eq:YK_highT_relation}
		F\approx\dfrac{1}{d_A^2P}.
	\end{equation}
	
	Therefore, at finite $T$, the corrections to both the probability and fidelity relative to the original YK decoder are encoded in the thermal factors $\eta_{\beta}$ and $\xi_{\beta}$.
	This distinguishes our protocol from the conventional finite-temperature YK protocol in Ref.~\cite{PhysRevD.106.046011}: the same global thermal quantity $\langle\rho_{\beta}^{1/2}\rangle=\Tr\rho_{\beta}^{1/2}/d_B$ also appears, whereas the additional factor $\xi_{\beta}$ originates from the SWAP injection and enters the recovery observables through the postselection on the expelled qubits $AA'$.
	Moreover, the SWAP-specific contribution of $\xi_{\beta}$ to both the probability and fidelity is weighted by $\kappa$, which scales as $O(1/d_{D}^{2})$.
	The two protocols therefore coincide for large $d_{D}$ and separate when $d_{D}$ is small, as shown numerically in Sec.~\ref{subsec:IIIC}.
	
	\vspace{5pt}
	\noindent
	\textbf{Initial values.}
	For the partition used in our simulations, the $t=0$ postselection probability is
	\begin{equation}
		\label{eq:P_early_beta}
		P_{\beta}(0)=\dfrac{1}{d_A d_D}\left\|\operatorname{Tr}_{bD}\rho_{\beta}^{1/2}\right\|_{F}^{2}.
	\end{equation}
	See Appendix~\ref{appendix:B} for the detailed derivation.
	Thus $P_{\beta\to0}(0)=1$ in the limit $\beta\to0$.
	The conditional fidelity starts from the trivial EPR overlap,
	\begin{equation}
		\label{eq:F_early_beta}
		F_{\beta}(0)=\dfrac{1}{d_A^2},
	\end{equation}
	regardless of $T$.
	See Appendix~\ref{appendix:C} for the detailed derivation.
	
	As shown in Fig.~\ref{fig:T_dependence}, the analytic late-time estimates in \eqref{eq:P_sat_beta} and \eqref{eq:F_sat_beta}, and the early-time values in \eqref{eq:P_early_beta} and \eqref{eq:F_early_beta}, agree well with the corresponding numerical results over the full temperature range considered here.
	This implies that the protocol is well described by the uniform-spreading assumption \eqref{eq18} for the chosen Hamiltonian model.
	
	\vspace{5pt}
	\noindent
	\textbf{OTOC.}
	At infinite temperature, the ideal $P_{\beta\to0}(t)$ is directly related to the OTOC~\cite{Roberts2017, PhysRevX.9.011006}.
	A similar qualitative comparison can be formulated in our protocol with a thermally regularized OTOC.
	Appendix~\ref{appendix:B} shows that $P_{\beta}(t)$ is governed by an OTOC-type correlator between $D$ and a thermally dressed operator on the injected mode $b$. 
	Motivated by this structure, we use the $b$-$D$ thermal OTOC as a diagnostic of operator growth from the injection site $b$ into $D$.
	
	Let $\mathcal{P}_{b}$ denote a Pauli operator on the injected qubit $b\in B$ and let $\mathcal{P}_{D}$ denote a Pauli string on the subsystem $D$.
	Similar to the results in \cite{PhysRevX.9.011006}, we consider
	\begin{equation}
		\label{eq:Pb_PD_OTOC}
		C_{\beta}(t)
		=
		\frac{1}{d_{b}^{2}d_{D}^{2}}
		\sum_{\mathcal{P}_{b},\mathcal{P}_{D}}
		\operatorname{Tr}\!\left[
		\mathcal{P}_{b} \mathcal{P}_{D}(t)\rho_{\beta}^{1/2}
		\mathcal{P}_{b}\mathcal{P}_{D}(t)\rho_{\beta}^{1/2}
		\right],
	\end{equation}
	where $\mathcal{P}_{D}(t)=e^{iHt}\mathcal{P}_{D} e^{-iHt}$.
	Here the sum over $\mathcal{P}_{b}$ runs over $\{I, X, Y, Z\}$, while the sum over $\mathcal{P}_{D}$ runs over the complete Pauli-string basis $\{I, X, Y, Z\}^{\otimes N_{D}}$.
	
\begin{figure}[tb]
		\centering
		\includegraphics[width=0.7\linewidth]{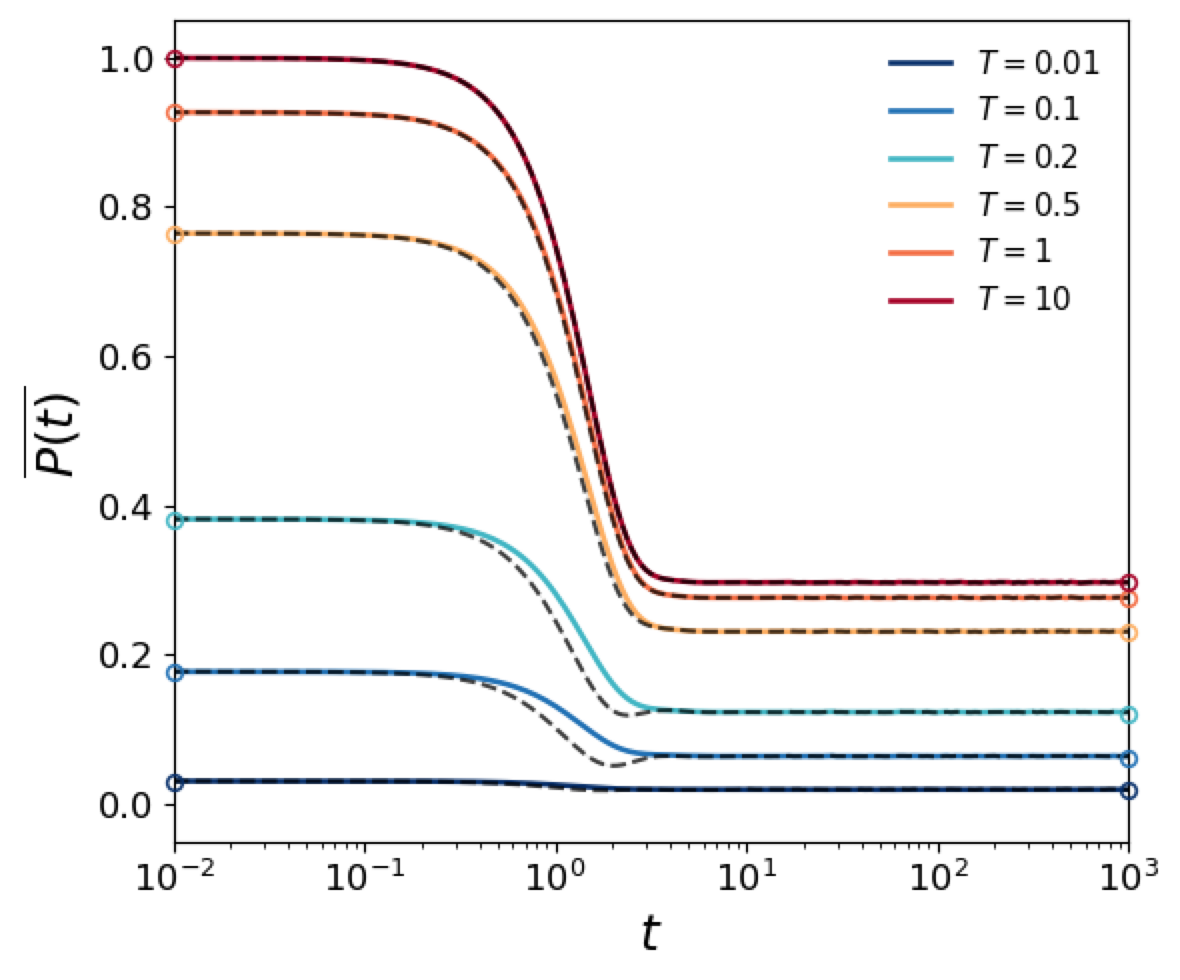}
		\caption{
			\label{fig:OTOC}
			Comparison between the Yoshida--Kitaev postselection probability $\overline{P_{\beta}(t)}$ (solid) and the thermal OTOC diagnostic (dashed), both averaged over 20 disorder samples, at $T=0.01,0.1,0.2,0.5,1,$ and $10$.
			We take $N_{A} = 1$, $N_{B} = 7$, $N_{C} = 5$, and $N_{D} = 2$ with an $N = 16$ SYK model.
			The qualitative agreement between them indicates that the decrease of $P_{\beta}(t)$ is consistent with the operator growth from the injected black-hole qubit $b$ into $D$.
		}
	\end{figure}

	For comparison with $\overline{P_{\beta}(t)}$, we plot $\overline{C_{\beta}(t)}$, taking $N_{C} = 5$ and $N_{D} = 2$.
	The resulting curve exhibits a time dependence qualitatively similar to $\overline{P_{\beta}(t)}$, as shown in Fig.~\ref{fig:OTOC}.
	The remaining difference at early times arises because the OTOC in \eqref{eq:Pb_PD_OTOC} replaces the full SWAP-block operator entering $P_{\beta}(t)$ by its Pauli-averaged local representative $\mathcal{P}_{b}$ on the injected site.
	Nevertheless, the agreement indicates that both the decay of $\overline{P_{\beta}(t)}$ and the growth of $\overline{F_{\beta}(t)}$ within the scrambling window are governed by the operator growth from the injected mode $b$ into the late-radiation subsystem $D$.
	This result is consistent with the conventional YK interpretation, where the postselection probability is closely related to scrambling diagnostics based on OTOCs.
	
	\section{\label{sec:quantum_computer_simulation}Implementation on a quantum computer}
	Our modified protocol admits a conventional HP recovery interpretation while also allowing us to study how recovery depends on explicit Hamiltonian dynamics.
	To study this experimentally, we implement the YK protocol on a quantum device.
	However, a dense SYK Hamiltonian results in a large circuit depth, reducing the reliability of the resulting data.
	We therefore consider sparse SYK Hamiltonians as hardware-feasible models.
	
	\subsection{Sparse SYK models}
	We briefly review two classes of sparse SYK models.
	
	\begin{figure*}[tb]
		\includegraphics[width=1.0\linewidth]{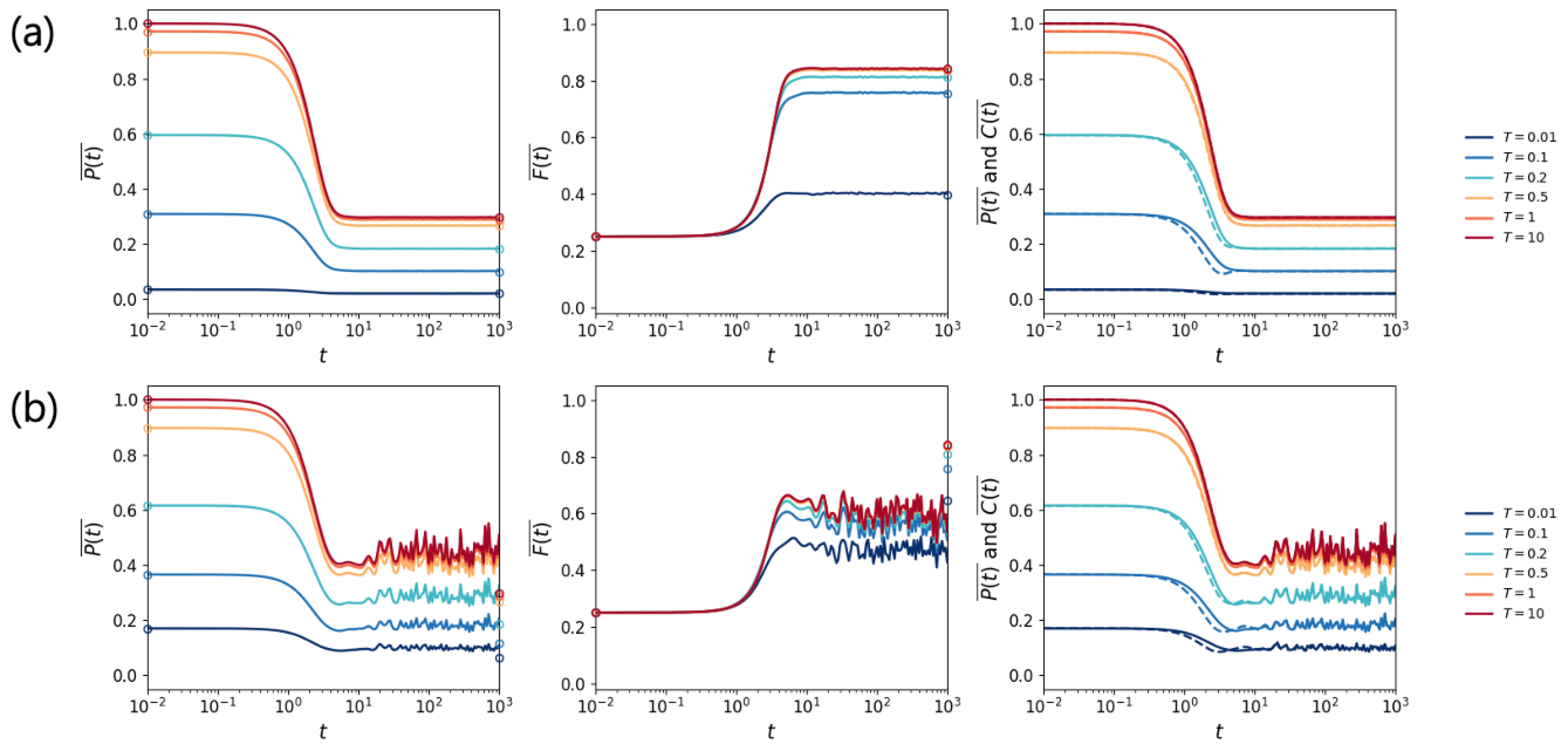}
		\caption{
			\label{fig:sparse}
			Effect of sparsification on the finite-temperature Yoshida--Kitaev protocol in the binary sparse $N = 16$ SYK model with $N_{A} = 1$, $N_{B} = 7$, $N_{C} = 5$, and $N_{D} = 2$.
			The postselection probability $\overline{P_{\beta}(t)}$, the conditional fidelity $\overline{F_{\beta}(t)}$, and the OTOC diagnostic are shown for (a) $K=100$ and (b) $K=6$ retained interaction terms, averaged over 20 disorder realizations.
			The case $K=100$ retains a stable recovery window and smooth OTOC saturation, while the strongly sparse case $K=6$ shows pronounced oscillations in both the decoding observables and the OTOC diagnostic.
			This indicates that stable recovery requires sufficiently strong operator growth in the present protocol.
		}
	\end{figure*}
	
	\vspace{5pt}
	\noindent
	\textit{1. Gaussian sparse SYK.}---
	The Gaussian sparse SYK Hamiltonian is given by
	\begin{align}
		\label{eq:sparse_1}
		H = i^{q/2}\sum_{1 \leq j_{1} < \cdots < j_{q} \leq N}\mathcal{J}_{j_{1}\cdots j_{q}}\chi_{j_{1}}\cdots\chi_{j_{q}},\\
		\label{eq:J}
		\mathcal{J}_{j_{1}\cdots j_{q}} = J_{j_{1}\cdots j_{q}}x_{j_{1}\cdots j_{q}},
	\end{align}
	where $x_{j_{1}\cdots j_{q}}\in\{0,1\}$ specifies whether the corresponding interaction term is retained.
	In the sparse ensemble, the nonzero Gaussian couplings have zero mean and variance
	\begin{equation}
		\label{eq:sparse_syk_variance}
		\left\langle J_{j_{1}\cdots j_{q}}^{2}\right\rangle
		=
		\frac{J^{2}(q-1)!}{pN^{q-1}}.
	\end{equation}
	Here $p$ denotes the retention probability, such that $x_{j_{1}\cdots j_{q}}=1$ with probability $p$.
	In our finite-size numerics, we fix the number of retained terms as $K=p\binom{N}{q}$ for a given $p$.\footnote{One can also consider a regular sparse ensemble, where each Majorana fermion appears in the same number of interaction terms and each interaction term contains the same number of fermions~\cite{xu2020sparsemodelquantumholography, Caceres2021}.}
	
	\vspace{5pt}
	\noindent
	\textit{2. Binary sparse SYK.}---
	Another sparsification scheme is the binary-coupling sparse SYK model~\cite{PhysRevB.107.L081103}.
	This model takes the coefficient in \eqref{eq:sparse_1} in the form
	\begin{equation}
		\mathcal{J}_{j_{1}\cdots j_{q}} = x_{j_{1}\cdots j_{q}}\,\eta_{j_{1}\cdots j_{q}}\,\dfrac{J}{\sqrt{K}},
		\quad
		\eta_{j_{1}\cdots j_{q}}\in\{+1,-1\},
	\end{equation}
	with the two signs chosen with equal probability, while the retained interaction terms are chosen by random pruning.
	For retained terms, all nonzero coefficients have the same magnitude $\mathcal{J} \equiv J/\sqrt{K}$.
	
	It is known that the binary sparse SYK model retains quantum-chaotic behavior under sparsification more robustly than the Gaussian sparse SYK model~\cite{PhysRevB.107.L081103}, as is often diagnosed by the spectral form factor (SFF) and the level-spacing distribution~\cite{PhysRevLett.110.084101,Cotler2017,Gharibyan2018,PhysRevD.98.086026,PhysRevB.107.L081103,Orman2025}.
	In black-hole-inspired quantum simulation, retaining these quantum-chaotic features is important for realizing scrambling dynamics.
	For this reason, we use the binary sparse SYK Hamiltonian to probe the recovery dynamics of the SYK model.
	
	\vspace{5pt}
	We numerically examine how sparsification affects the recovery dynamics.
	The recovery behavior remains robust as long as the sparse SYK Hamiltonian stays in the chaotic regime.
	When the Hamiltonian becomes too sparse, both $\overline{P_{\beta}(t)}$ and $\overline{F_{\beta}(t)}$ develop strong late-time oscillations, with reduced $\overline{F_{\beta}(t)}$.
	For instance, Fig.~\ref{fig:sparse} shows $\overline{P_{\beta}(t)}$, $\overline{F_{\beta}(t)}$, and $\overline{C_{\beta}(t)}$ for binary sparse $N = 16$ SYK Hamiltonians with $K=100$ ($p \approx 0.055$) and $K=6$ ($p \approx 0.003$).
	In the $K=6$ case, $\overline{F_{\beta}(t)}$ is strongly suppressed and exhibits pronounced late-time oscillations.
	Indeed, the late-time limits of both $\overline{P_{\beta}(t)}$ and $\overline{F_{\beta}(t)}$ largely deviate from the analytic expectations.
	
	One notable point is that, even in the strongly sparse case, the oscillatory behavior of $\overline{P_{\beta}(t)}$ follows $\overline{C_{\beta}(t)}$.
	This suggests that the OTOC continues to track the postselection probability even in the strongly sparse regime, indicating that the loss of recovery there reflects a loss of stable, ergodic operator spreading.
	
	\subsection{Circuit construction}
	For the hardware implementation, we use the even-parity sector of an $N=8$ SYK Hamiltonian, represented by $N_{B}=3$ qubits.
	Taking $N_{A}=N_{R}=N_{C}=1$ and $N_{D}=2$, the full circuit therefore uses $10$ qubits.
	As in the numerical analysis, the diary is injected into the first qubit in $B$, with the corresponding mirror qubit used in $B'$.
	We choose $T = 0.1$ and $10$ to probe the low- and high-temperature regimes, respectively.
	The full Hamiltonian is a binary sparse $N = 8$ SYK Hamiltonian with $K=10$ ($p\approx0.14$), given by
	\begin{widetext}
		\begin{equation}
			\label{eq:syk_hamiltonian_N8}
			\begin{aligned}
				\dfrac{H}{\mathcal{J}} &= \chi_{2}^{}\chi_{3}^{}\chi_{4}^{}\chi_{6}^{} + \chi_{1}^{}\chi_{2}^{}\chi_{3}^{}\chi_{8}^{} + \chi_{1}^{}\chi_{3}^{}\chi_{5}^{}\chi_{7}^{} + \chi_{4}^{}\chi_{5}^{}\chi_{6}^{}\chi_{8}^{} + \chi_{3}^{}\chi_{5}^{}\chi_{6}^{}\chi_{8}^{} - \chi_{1}^{}\chi_{2}^{}\chi_{4}^{}\chi_{7}^{} - \chi_{1}^{}\chi_{4}^{}\chi_{7}^{}\chi_{8}^{} - \chi_{3}^{}\chi_{4}^{}\chi_{5}^{}\chi_{6}^{}\\
				&\quad - \chi_{3}^{}\chi_{6}^{}\chi_{7}^{}\chi_{8}^{} - \chi_{1}^{}\chi_{2}^{}\chi_{4}^{}\chi_{6}^{},
			\end{aligned}
		\end{equation}
	\end{widetext}
	where $\mathcal{J} = \frac{1}{\sqrt{5}}$.
	This Hamiltonian is chosen such that (i) quantum chaos is preserved~\cite{byun2026quantumsimulationtraversablewormholeinspiredquantum} and (ii) the single-step Lie--Trotterized dynamics closely reproduces the exact time evolution over the time interval $t\in[0,6]$.
	The latter criterion allows us to avoid the deeper quantum circuits required by higher-order approximation or additional Trotter steps.
	
	
	We focus on the time window $t\in[0,6]$, within which $P_{\beta}(t)$ and $F_{\beta}(t)$ develop toward saturation and exhibit clear temperature dependence.
	At later times, the hardware-scale system shows finite-size fluctuations unrelated to sparsification, so the experimentally relevant decoding dynamics is probed within this interval.
	
	\begin{figure}[t]
		\centering
		\includegraphics[width=\linewidth]{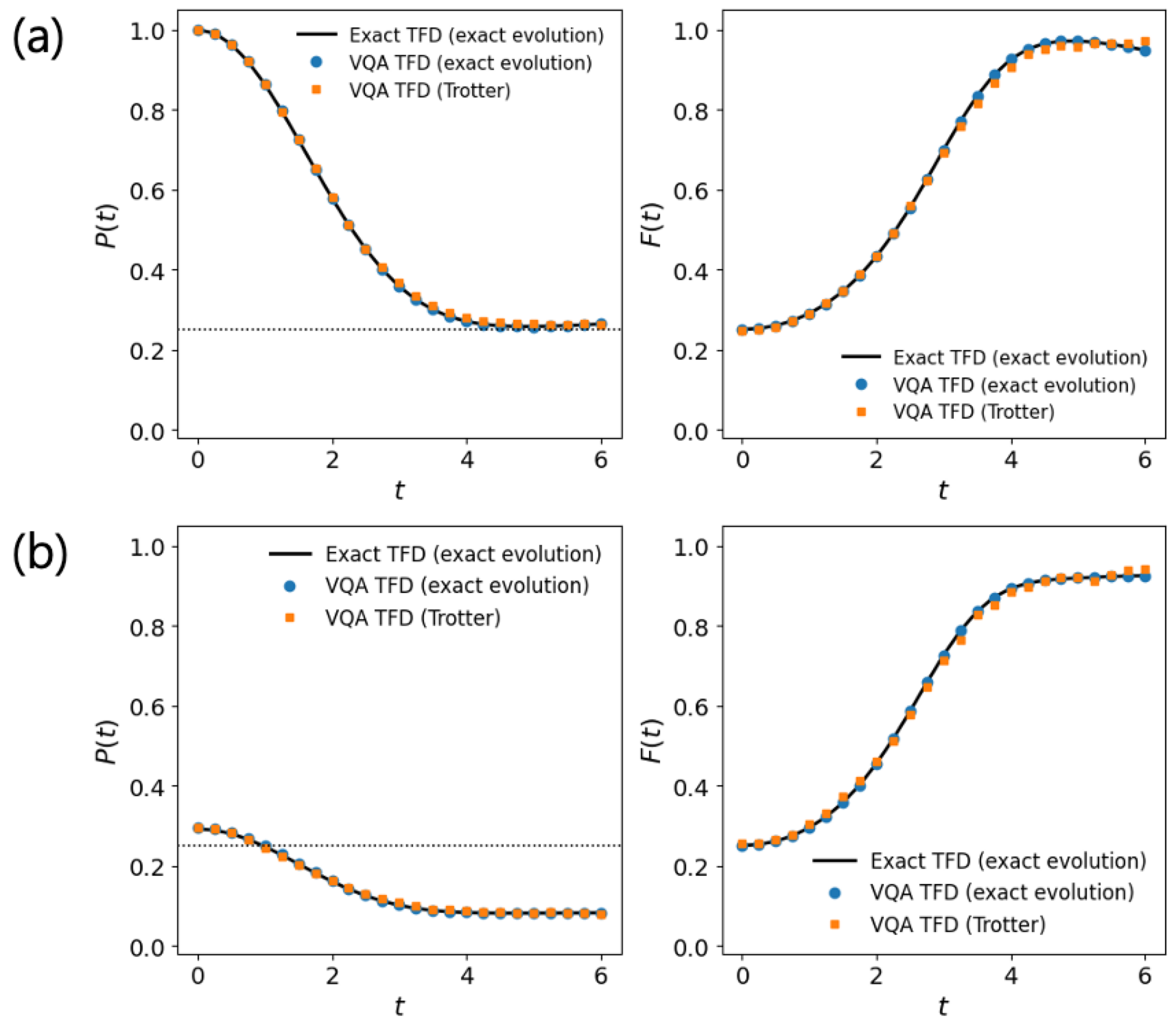}
		\caption{
			\label{fig:trotter}
			The postselection probability $P(t)$ and the conditional fidelity $F(t)$ at (a) $T = 10$ and (b) $T = 0.1$, obtained by exact time evolution from the exact TFD statevector (black) and from the variationally prepared TFD state (blue).
			These are compared with single-step first-order Lie--Trotter evolution from the variationally prepared TFD state (orange).
			The dotted horizontal lines indicate the common reference value $1/d_A^2=1/4$.
			The Trotterized dynamics from the VQA state accurately reproduces the exact time evolution within $t\in[0,6]$.
		}
	\end{figure}
	
	For the circuit construction, we first prepare the TFD state using a variational quantum algorithm (VQA)~\cite{Kandala2017,PhysRevLett.123.220502,PhysRevA.104.012427}.
	We construct the exact TFD statevector $\ket{\text{TFD}}$ classically from the chosen Hamiltonian \eqref{eq:syk_hamiltonian_N8}.
	This exact state is used as the target state for a variational quantum circuit acting on the $BB'$ register.
	The variational circuit starts with Hadamard gates applied to all qubits in $BB'$, followed by repeated layers of $R_x$ and $R_z$ rotations and linear entangling gates.
	Denoting the state prepared by the variational circuit as $\ket{\psi_{\text{VQA}}(\boldsymbol{\theta})}$ with variational parameters $\boldsymbol{\theta}$, we optimize $\boldsymbol{\theta}$ by maximizing the fidelity with the exact TFD state,
	\begin{equation}
		\label{eq:vqa_fidelity}
		\boldsymbol{\theta}^{\star} = \argmax_{\boldsymbol{\theta}} \, |\innerproduct{\psi_{\text{VQA}}(\boldsymbol{\theta})}{\text{TFD}}|^{2},
	\end{equation}
	and the optimized state $\ket{\psi_{\text{VQA}}(\boldsymbol{\theta}^{\star})}$ is used as the prepared TFD state.
	For the chosen Hamiltonian, we achieve a fidelity of $99.99\%$ at $T=10$ using an ansatz circuit comprising 77 single-qubit gates and 15 CZ gates.
	At $T=0.1$, we achieve a fidelity of $99.92\%$ using 115 single-qubit gates, 24 CZ gates, and one $R_{zz}$ gate.
	Figure~\ref{fig:trotter} shows that the variationally prepared TFD states closely reproduce $P_{\beta}(t)$ and $F_{\beta}(t)$ obtained from the exact TFD statevector.

	After applying the SWAP gates, the real-time evolution is implemented by a Trotterization~\cite{82edc856-4d85-3b98-9b0d-ad55bb9315f6,Seth1996} over the chosen time interval $t\in[0,6]$.
	We use a single-step Lie Trotterization, and the resulting $P_{\beta}(t)$ and $F_{\beta}(t)$ are shown in Fig.~\ref{fig:trotter}.
	We find that the Trotterization closely reproduces the behavior of $P_{\beta}(t)$ and $F_{\beta}(t)$ for each temperature within the chosen time range, so we adopt this Trotterization for our experiment.\footnote{Higher-order product formulas and a larger number of Trotter steps may be required to obtain consistent recovery behavior at later times $t > 6$.}
	
	In the experiment, the probabilistic YK decoder is implemented directly by Bell measurements.
	For each pair of qubits in $A$ and $A'$, and in $D$ and $D'$, we apply a CNOT gate followed by a Hadamard gate, such that the pre-measurement state becomes
	\begin{equation}
		H_{A}\operatorname{CNOT}_{A\to A'}
		\prod_{k=1}^{N_D}
		\left(
		H_{D_k}\operatorname{CNOT}_{D_k\to D'_k}
		\right)
		|\widetilde{\mathrm{out}}(t)\rangle.
	\end{equation}
	We then measure this state in the $Z$ basis.
	In this setup, the postselection succeeds only when the measurement outcome corresponds to $\ket{0}^{\otimes N_{AA'DD'}}$ where $N_{AA'DD'} = N_{A}+N_{A'}+N_{D}+N_{D'}$.
	The probability $P_{\beta}(t)$ is therefore obtained from the raw counts as
	\begin{equation}
		P_{\beta}(t) = \frac{N_{AA'=00,\,DD'=00\cdots0}}{N_{\mathrm{shot}}},
	\end{equation}
	where $N_{\mathrm{shot}}$ denotes the total number of measurement shots at each time point.
	To measure $F_{\beta}(t)$, we also perform a Bell-basis measurement on $RR'$ in the same circuit.
	Using \eqref{eq23}, the fidelity is obtained as
	\begin{equation}
		F_{\beta}(t)
		=
		\frac{N_{RR'=00,\,AA'=00,\,DD'=00\cdots0}}{N_{AA'=00,\,DD'=00\cdots0}}.
	\end{equation}
	In our experiment, we take $N_{\rm shot} = 10{,}000$.
	
	\begin{figure}[tb]
		\centering
		\includegraphics[width=0.8\linewidth]{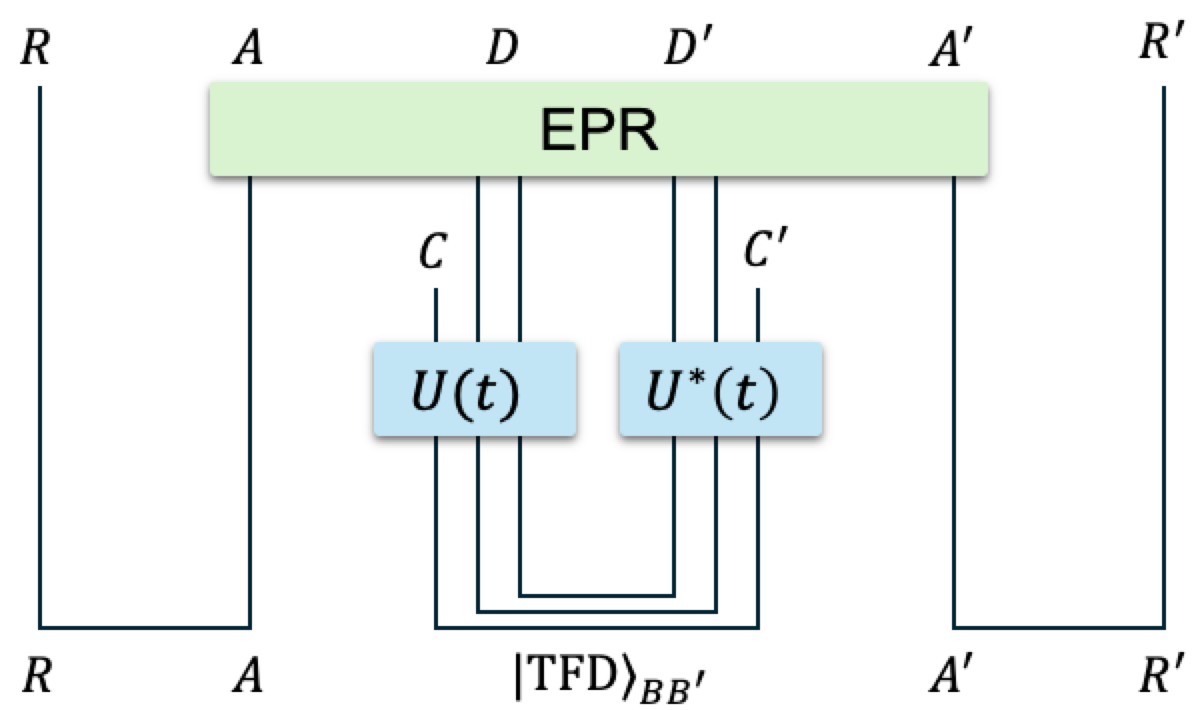}
		\caption{
			\label{fig:mitigation_circuit}
			The reference circuit used to estimate noise-induced suppression for $P_{\beta}(t)$ and $F_{\beta}(t)$.
			The TFD preparation and the forward and conjugated time evolutions are kept identical to the target Yoshida--Kitaev circuit, while the SWAP injections are removed.
			In the ideal noiseless circuit, the reference probability and fidelity are independent of time because $U_B(t)U_{B'}^{*}(t)$ leaves the TFD state invariant.
		}
	\end{figure}
	
	\subsection{Error mitigation}
	To estimate the effect of hardware noise, we use a reference circuit in which the same TFD preparation and the same time evolution are applied, while the SWAP injections $S_{Ab}$ and $S_{A'b'}$ are removed.
	This provides a reference response without inserting the diary information into the black-hole system.
	The circuit construction is shown in Fig.~\ref{fig:mitigation_circuit}.
	
	In the ideal noiseless simulation, the reference postselection probability is independent of $t$; with the SWAP gates removed, the forward and conjugated evolutions $U_{B}(t)U_{B'}^{*}(t)$ leave the TFD state invariant.
	Therefore, the reference probability is
	\begin{equation}
		\label{eq:ctrl_P}
		P_{\mathrm{ref}}(t)=\bra{\mathrm{in}}\Pi_{AA'}\Pi_{DD'}\ket{\mathrm{in}} = \dfrac{1}{d_{A}^{2}d_{D}}\|\Tr_{D}\rho_{\beta}^{1/2}\|_{F}^{2}.
	\end{equation}
	This time independence makes the reference circuit a clean baseline for hardware-induced errors, since any $t$ dependence in the measured $P_{\mathrm{ref}}(t)$ may arise from noise.
	In this sense, the construction supplies a natural, protocol-adapted reference for error mitigation, in the same spirit as the strategy used to measure OTOCs in \cite{doi:10.1126/science.abg5029}.
	Therefore, hardware errors in the TFD preparation, time evolution, and measurement are expected to manifest as a suppression of \eqref{eq:ctrl_P}.
	This suppression can be quantified by defining the mitigation factor
	\begin{equation}
		\label{eq39}
		M_{P}(t)=\frac{P_{\mathrm{ref}}^{\mathrm{ideal}}(t)}{P_{\mathrm{ref}}^{\mathrm{noisy}}(t)},
	\end{equation}
	where $P_{\mathrm{ref}}^{\mathrm{ideal}}(t)$ and $P_{\mathrm{ref}}^{\mathrm{noisy}}(t)$ are the postselection probability in the noiseless reference circuit and that obtained from the noisy hardware experiment.
	Then, we apply \eqref{eq39} to the experimental data of $P_{\beta}^{\mathrm{noisy}}(t)$, so the error-mitigated probability is
	\begin{equation}
		P_{\beta}^{\rm mit}(t) = M_{P}(t)P_{\beta}^{\mathrm{noisy}}(t).
	\end{equation}
	
	The SWAP-removed reference circuit can also be used to mitigate the conditional fidelity using an approach similar to that developed in Ref.~\cite{PhysRevLett.127.270502}.
	To this end, we approximate the effective hardware noise acting on $RR'$ by a depolarizing channel,
	\begin{equation}
		\label{eq:depol_model}
		\rho_{RR'}^{\mathrm{noisy}}(t)
		=
		\lambda(t)\,\rho_{RR'}^{\mathrm{ideal}}(t)
		+\bigl[1-\lambda(t)\bigr]\frac{I_{RR'}}{d_A^{2}},
	\end{equation}
	with a single suppression parameter $0<\lambda(t)\le 1$. 
	Since $F_{\beta}$ is linear in $\rho_{RR'}$, it is affine in $\lambda$,
	\begin{equation}
		\label{eq:affine_F}
		F_{\beta}^{\mathrm{noisy}}(t)
		=
		\lambda(t)\,F_{\beta}^{\mathrm{ideal}}(t)
		+\bigl[1-\lambda(t)\bigr]\,F_{\infty},
	\end{equation}
	where the fixed point $F_{\infty} = 1/d_{A}^{2}$ is the EPR overlap of the maximally mixed state. 
	Now we determine $\lambda(t)$ from the reference circuit. 
	Since the reference and target circuits differ only by the two SWAP gates, we assume, as in Ref.~\cite{PhysRevLett.127.270502}, that they are subject to approximately the same effective suppression parameter $\lambda(t)$.
	Crucially, since the reference dynamics leaves the $RR'$ EPR pair intact, its ideal conditional fidelity is $F_{\mathrm{ref}}^{\mathrm{ideal}}(t)=1$, so that \eqref{eq:affine_F} applied to the reference gives
	\begin{equation}
		\label{eq:lambda}
		\lambda(t)
		=
		\frac{F_{\mathrm{ref}}^{\mathrm{noisy}}(t)-F_{\infty}}
		{1-F_{\infty}}.
	\end{equation}
	Inverting \eqref{eq:affine_F} for the target circuit, the error-mitigated fidelity is
	\begin{equation}
		\label{eq:F_mitigated}
		F_{\beta}^{\mathrm{mit}}(t)
		=
		F_{\infty}
		+\frac{F_{\beta}^{\mathrm{noisy}}(t)-F_{\infty}}{\lambda(t)}.
	\end{equation}

	\begin{figure*}[tb]
		\centering
		\includegraphics[width=0.8\linewidth]{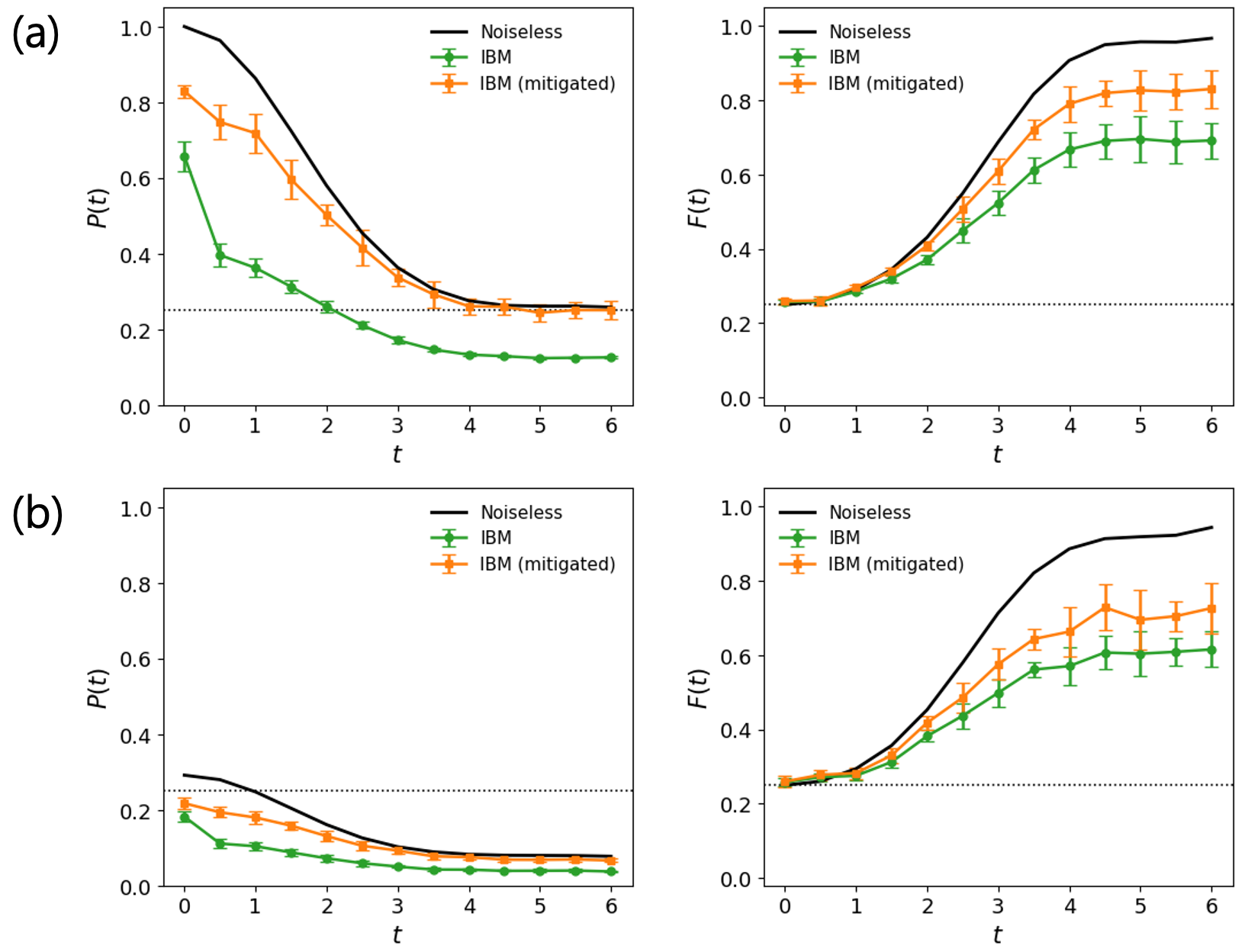}
		\caption{
			\label{fig:yk_quantum_computer}
			Quantum-computer implementation of the SWAP-injected Yoshida--Kitaev decoder on \texttt{ibm\_marrakesh} at (a) $T = 10$ and (b) $T = 0.1$.
			The left and right panels show the postselection probability $P(t)$ and the conditional fidelity $F(t)$, respectively.
			The circuit uses $N_{A} = 1$, $N_{B} = 3$, $N_{C} = 1$, and $N_{D} = 2$, with an $N=8$ binary sparse SYK Hamiltonian with $K=10$ retained terms, and each time point is measured with $10{,}000$ shots.
			The raw hardware data (green), the reference-circuit-mitigated results $P_{\beta}^{\rm mit}(t)$ and $F_{\beta}^{\rm mit}(t)$ (orange), and the noiseless results (black) are shown.
			Markers and error bars denote the mean and standard deviation, respectively, over five independent repetitions of the full experiment.
			The dotted horizontal lines indicate the common reference value $1/d_A^2=1/4$.
			In the scrambled regime, the reference-circuit-based mitigation substantially restores $P_{\beta}(t)$ toward the noiseless case, and partially restores $F_{\beta}(t)$ with a residual gap remaining.
		}
	\end{figure*}
	
	\subsection{Hardware results}
	We implement the experiment on the IBM superconducting processor \texttt{ibm\_marrakesh}.
	The 13 time points are taken in the interval $t\in[0,6]$ for each temperature.
	Each point is measured with $10{,}000$ shots, and we repeat the full experiment five times to obtain the standard deviation.
	Except for $t=0$, the circuit at each time point contains 359 one-qubit gates and 104 two-qubit gates, with a total depth of 144, at $T=10$, and 397 one-qubit gates and 114 two-qubit gates, with a total depth of 156, at $T=0.1$.
	
	Figure~\ref{fig:yk_quantum_computer} shows the hardware results together with the corresponding noiseless results obtained by applying the single-step Lie--Trotterized evolution to the variationally prepared statevector $\ket{\psi_{\mathrm{VQA}}(\boldsymbol{\theta}^{\star})}$.
	Device noise suppresses $P_{\beta}(t)$ and degrades $F_{\beta}(t)$ for both $T = 10$ and $T = 0.1$.
	Nevertheless, the hardware data retain the qualitative recovery dynamics, exhibiting a decreasing postselection probability together with an increasing conditional fidelity.
	The lower postselection probability at lower temperature is also consistent with the noiseless numerics.
	As in Ref.~\cite{Landsman2019}, the simultaneous increase of the fidelity indicates that the decrease in probability is not caused solely by device noise, but also reflects the intrinsic decoding dynamics of the YK protocol.
	
	The error-mitigated hardware results are also shown in Fig.~\ref{fig:yk_quantum_computer}.
	In the scrambled regime, the reference-circuit-based mitigation substantially restores $P_{\beta}(t)$ toward the noiseless case through the factor $M_{P}(t)$, while the depolarizing-channel-based mitigation partially restores $F_{\beta}(t)$.
	The mitigated results imply that the two quantities are affected by noise in different ways.
	The probability suppression is modeled as predominantly multiplicative as in \eqref{eq39}, so that the single factor $M_{P}(t)$ restores it close to the noiseless case.
	For the fidelity, however, such a multiplicative correction does not apply, and the depolarizing model \eqref{eq:F_mitigated} partially recovers it.
	A more appropriate mitigation of $F_{\beta}(t)$ would require techniques beyond the depolarization model, such as probabilistic error cancellation or zero-noise extrapolation applied to the joint and marginal count distributions~\cite{PhysRevX.7.021050,PhysRevLett.119.180509}; this is left for future work.

	\section{\label{sec:discussion}Discussion}
	We have studied a finite-temperature YK decoding protocol with explicit SYK Hamiltonian dynamics.
	Our results show that finite-temperature Hayden--Preskill recovery can be realized by SWAP injection without enlarging the black-hole Hilbert space or replacing the Hamiltonian by an extended one.
	
	In the SWAP-injected construction, the diary is inserted into an existing qubit of the black-hole system, so that the same microscopic Hamiltonian defines the TFD state and generates the subsequent scrambling dynamics.
	The register $A$ after the SWAP is then interpreted as the initial radiation, consisting of the pre-existing black-hole qubit expelled at the diary-injection step.
	
	At finite temperature, the late-time saturation values of the postselection probability and conditional fidelity admit analytic estimates under the uniform-spreading assumption.
	These are represented in terms of two thermal factors $\eta_{\beta}$ and $\xi_{\beta}$, which are determined by $\rho_{\beta}^{1/2}$ and its reduced operators.
	These factors also distinguish the SWAP-injected construction from the conventional appended decoder: when the late-radiation subsystem is small, the conditional fidelity acquires an additional temperature dependence and departs from the appended YK result, whereas the two protocols coincide when the late-radiation subsystem is large.
	The comparison with the OTOC further shows that the recovery dynamics follows operator growth from the injected mode into the late radiation.
	
	For the hardware implementation, we examined the effect of Hamiltonian sparsification.
	The decoding behavior remains stable when the sparse SYK Hamiltonian retains sufficient chaotic dynamics, while excessive sparsification produces oscillatory behavior and reduced fidelity.
	This makes binary sparse SYK a useful model for reducing circuit complexity while preserving the recovery signal in the present protocol.
	
	Finally, we implemented the finite-temperature YK decoder on an IBM superconducting quantum computer at low and high temperatures.
	The noisy data preserve the qualitative trend of a decreasing postselection probability together with an enhanced conditional fidelity.
	The expected temperature dependence is also observed in the hardware results.
	Moreover, the reference-circuit-based mitigation substantially restores the postselection probability and partially restores the conditional fidelity.
	
	Our work therefore provides a well-defined, single-Hamiltonian, and hardware-compatible realization of black-hole-inspired quantum information recovery and demonstrates its implementation on a noisy intermediate-scale quantum (NISQ) device~\cite{Preskill2018quantumcomputingin}.
	Future directions include deterministic YK decoding, improved mitigation methods for the conditional fidelity, and injection operators beyond the ordinary SWAP.
	Extending the protocol to other scrambling-based recovery schemes, such as bidirectional teleportation, is also a promising direction~\cite{vikram2026bidirectionalteleportationusingscrambling, sun2026postselectionprobabilityfidelitybidirectional}.
	More broadly, SWAP-based injection may also be useful in other black-hole-inspired quantum circuits, for example, those designed to probe entanglement dynamics.
	
	
	\section*{Acknowledgements}
	This work was supported by the Basic Science Research Program through the National Research Foundation of Korea(NRF) funded by the Ministry of Science, ICT \& Future Planning(NRF-2021R1A2C1006791), the Korea government(MSIT)(RS-2025-02311201), (RS-2024-00445164) and the framework of international cooperation program managed by the NRF of Korea(RS-2025-02307394), the Creation of the Quantum Information Science R\&D Ecosystem(Grant No. RS-2023-NR068116) through the National Research Foundation of Korea(NRF) funded by the Korean government(Ministry of Science and ICT). 
	This research was also supported by GIST research fund (Future leading Specialized Research Project, 2026, and the Regional Innovation System \& Education(RISE) program through the(Gwangju RISE Center), funded by the Ministry of Education(MOE) and the(Gwangju Metropolitan City), Republic of Korea(2025-RISE-05-001).
	J.B. and K.C. were supported by the Ministry of Science, ICT and Future Planning (MSIP) by the National Research Foundation of Korea (RS-2024-00432214);  the Korean ARPA-H Project through the Korea Health Industry Development Institute (KHIDI), funded by the Ministry of Health \& Welfare, Republic of Korea (RS-2025-25456722). We acknowledge the Yonsei University Quantum Computing Project Group for providing support and access to the Quantum System One (Eagle Processor), which is operated at Yonsei University.

	\appendix
	\section*{Appendices}

	\section{\label{appendix:B}Postselection probability at finite temperature}
	In this appendix, we derive the initial value and the late-time estimate of the postselection probability.
	In the full initial state \eqref{eq:initial_state} in the YK protocol, we denote
	\begin{equation}
		\begin{aligned}
			\ket{\rm EPR}_{RA} &= \frac{1}{\sqrt{d_A}}\sum_{a=0}^{d_A-1}\ket{a}_{R}\ket{a}_{A}\\
			\ket{\rm EPR}_{R'A'} &= \frac{1}{\sqrt{d_{A'}}}\sum_{r=0}^{d_{A'}-1}\ket{r}_{R'}\ket{r}_{A'}
		\end{aligned}
	\end{equation}
	We restrict to the case $d_A=d_{A'}=2$, but the arguments below can be generalized to the case $d_{A} > 2$.
	We decompose the black-hole Hilbert spaces as $B=b\bar{B}$ and $B'=b'\bar{B}'$, where $b$ and $b'$ are the qubits swapped with $\ket{a}$ on $A$ and $\ket{r}$ on $A'$, respectively.
	For the SYK realization, this decomposition is understood within the parity-reduced qubit representation discussed in Sec.~\ref{subsec:syk_model}.
	The SWAP gates can then be written as
	\begin{equation}
		\begin{aligned}
			S_{Ab} &= \sum_{a,a'\in\{0,1\}}\outerproduct{a'}{a}_{A}\otimes \mathcal{W}_{b}^{(a'a)},\\
			S_{A'b'} &= \sum_{r,r'\in\{0,1\}}\outerproduct{r'}{r}_{A'}\otimes \mathcal{V}_{b'}^{(r'r)}.
		\end{aligned}
	\end{equation}
	Here $\mathcal{W}_{b}^{(a'a)}$ and $\mathcal{V}_{b'}^{(r'r)}$ are operators acting on $B=b\bar{B}$ and $B'=b'\bar{B}'$, respectively, defined by
	\begin{equation}
		\begin{aligned}
			\mathcal{W}_{b}^{(a'a)} &= \outerproduct{a}{a'}_{b}\otimes I_{\bar{B}},\\
			\mathcal{V}_{b'}^{(r'r)} &= \outerproduct{r}{r'}_{b'}\otimes I_{\bar{B}'}.
		\end{aligned}
	\end{equation}
	The identity operators $I_{\bar{B}}$ and $I_{\bar{B}'}$ act on the degrees of freedom that do not participate in the SWAP.
	With these definitions, $S_{Ab}$ acts as
	\begin{equation}
		S_{Ab}\ket{a}_{A}\ket{x}_{b}\ket{\eta}_{\bar B}=\ket{x}_{A}\ket{a}_{b}\ket{\eta}_{\bar B},
	\end{equation}
	and similarly for $S_{A'b'}$.
	After applying these SWAP gates and the time-evolution operators $U_{B}(t)=e^{-iHt}$ and $U_{B'}^{\ast}(t)$ on $B$ and $B'$, respectively, the final state becomes
	\begin{widetext}
		\begin{equation}
			\label{eqB4}
			|\widetilde{\text{out}}(t)\rangle=\frac{1}{d_A}\sum_{a,a',r,r'}\ket{aa'}_{RA}\otimes\left[\mathcal{W}_{b}^{(a'a)}(t)\otimes \widetilde{\mathcal{V}}_{b'}^{(r'r)}(t)\right]\ket{\mathrm{TFD}}_{BB'}\otimes\ket{rr'}_{R'A'}.
		\end{equation}
	\end{widetext}
	Here, we have defined
	\begin{equation}
		\begin{aligned}
			\label{eqB5}
			\mathcal{W}_{b}^{(a'a)}(t) &= U_{B}(t)\mathcal{W}_{b}^{(a'a)}U_{B}^{\dagger}(t),\\
			\widetilde{\mathcal{V}}_{b'}^{(r'r)}(t) &= U_{B'}^{\ast}(t)\mathcal{V}_{b'}^{(r'r)}U_{B'}^{T}(t).
		\end{aligned}
	\end{equation}
	and used the fact that $U_{B}^{\dagger}(t)U_{B'}^{T}(t)$ leaves the TFD state invariant.
	
	Since $A$ after the SWAP contains the qubit expelled as radiation at the injection step, the EPR projection is performed on $AA'$ as well as $DD'$, as described in the main text.
	We therefore consider the postselection probability \eqref{eq:postselection_probability}, where the projective measurement $\Pi_{DD'}$ admits the Pauli-string expansion~\cite{PhysRevA.105.032435, PhysRevX.12.031013}
	\begin{equation}
		\label{eqA7}
		\Pi_{DD'}=\frac{1}{d_D^2}\sum_{\alpha}\mathcal{P}_{\alpha}\otimes\mathcal{P}_{\alpha}^{\ast}.
	\end{equation}
	Here $\{\mathcal{P}_{\alpha}\}$ is the complete Pauli-string basis on $D$, with $\mathcal{P}_{\alpha}\in\{I, X, Y, Z\}^{\otimes N_{D}}$, and $\{\mathcal{P}_{\alpha}^{\ast}\}$ denotes the corresponding complex-conjugate basis on $D'$.
	The normalization is $\Tr(\mathcal{P}_{\alpha}^{\dagger}\mathcal{P}_{\beta})=d_D\delta_{\alpha\beta}$.
	When inserted into operators on $B=CD$, $\mathcal{P}_{\alpha}$ is understood as $I_C\otimes\mathcal{P}_{\alpha}$.
	The measurement $\Pi_{AA'}$ is also the EPR projector.
	To proceed, we define
	\begin{equation}
		\label{eq:B_Nblock}
		\begin{aligned}
			\mathcal{N}^{(ar)} &:= \sum_{n=0}^{d_A-1}\mathcal{W}_{b}^{(na)}\,\rho_{\beta}^{1/2}\,\mathcal{W}_{b}^{(nr)\dagger},\\
			\mathcal{N}^{(ar)}(t)&=U(t)\mathcal{N}^{(ar)}U^{\dagger}(t).
		\end{aligned}
	\end{equation}
	We also identify $S_{A'b'}$ as the mirror copy of $S_{Ab}$, so that $\mathcal{V}_{b'}^{(r'r)}=\mathcal{W}_{b}^{(r'r)}$ and the time-evolved blocks obey the corresponding relation $\widetilde{\mathcal{V}}_{b'}^{(r'r)}(t)=(\mathcal{W}_{b}^{(r'r)}(t))^*$.
	Then, using the identity
	\begin{equation}
		\bra{\mathrm{TFD}}O_B\otimes O_{B'}\ket{\mathrm{TFD}}=\Tr\big[\rho_{\beta}^{1/2}O_B\rho_{\beta}^{1/2}O_{B'}^{T}\big]
	\end{equation}
	for some operators $O_{B}$ and $O_{B'}$ on $\mathcal{H}_{B}$ and $\mathcal{H}_{B'}$, respectively, and the fact that $[\rho_{\beta},U(t)]=0$, the postselection probability becomes
	\begin{equation}
		\label{eqB9}
		P_{\beta}(t)=\frac{1}{d_A^3 d_D^2}\sum_{a,r,\alpha}\Tr\!\left[\mathcal{P}_{\alpha}\mathcal{N}^{(ar)}(t)\mathcal{P}_{\alpha}\mathcal{N}^{(ar)\dagger}(t)\right].
	\end{equation}
	Thus, each postselection-probability element is again written as an OTOC-type correlator, now with the injected diary mode summed.
	We then use the Pauli completeness relation~\cite{PhysRevA.80.012304, Hosur2016}: for an operator $X$ on $\mathcal{H}_{C}\otimes\mathcal{H}_{D}$, we have
	\begin{equation}
		\label{eqB10}
		\sum_{\alpha}\mathcal{P}_{\alpha}X\mathcal{P}_{\alpha}=\Tr_D X\otimes d_D I_D.
	\end{equation}
	Applying this identity to \eqref{eqB9}, we find
	\begin{equation}
		\label{eq:B_P_reduced}
		P_{\beta}(t)=\frac{1}{d_A^3 d_D}\sum_{a,r}\left\|\Tr_D\mathcal{N}^{(ar)}(t)\right\|_F^2,
	\end{equation}
	where $\|\cdots\|_F$ denotes the Frobenius norm, $\|\mathcal{O}\|_F^2=\Tr(\mathcal{O}^{\dagger}\mathcal{O})$.
	
	We now estimate the late-time probability at finite temperature.
	Because the SYK Hamiltonian is an ensemble-defined model, we take the disorder average to characterize the typical late-time asymptotic value rather than sample-dependent fluctuations.
	The disorder-averaged postselection probability can be written as
	\begin{equation}
		\label{eqB12}
		\overline{P_{\beta}(t)}=\dfrac{1}{d_{A}^{3}d_{D}}\sum_{a,r}\overline{\|\Tr_{D}\mathcal{N}^{(ar)}(t)\|_{F}^{2}}.
	\end{equation}
	We decompose each $\mathcal{N}^{(ar)}$ into identity and traceless parts as
	\begin{equation}
		\label{eqB13}
		\mathcal{N}^{(ar)}=\mathcal{N}^{(ar)}_{I}+\mathcal{N}^{(ar)}_{0},
		\quad
		\mathcal{N}_{I}^{(ar)} = \dfrac{\Tr\mathcal{N}^{(ar)}}{d_{B}}I_{B},
	\end{equation}
	satisfying $\Tr\mathcal{N}^{(ar)}_{0}=0$.
	We also rewrite $\rho_{\beta}^{1/2}$ as
	\begin{equation}
		\label{eq:C5}
		\rho_{\beta}^{1/2}=\sum_{i,j=0}^{d_A-1}\outerproduct{i}{j}_{b}\otimes R_{ij},
		\quad
		R_{ij}={}_{b}\!\bra{i}\rho_{\beta}^{1/2}\ket{j}_{b}.
	\end{equation}
	Then, $\mathcal{N}^{(ar)}$ in \eqref{eq:B_Nblock} becomes
	\begin{equation}
		\label{eq:C_Nexplicit}
		\mathcal{N}^{(ar)}=\outerproduct{a}{r}_{b}\otimes \Tr_{b}\rho_{\beta}^{1/2} = \outerproduct{a}{r}_{b}\otimes\sum_{i=0}^{d_A-1}R_{ii}.
	\end{equation}
	The identity block $\mathcal{N}^{(ar)}_{I}\propto I_{B}$ is invariant under $U(t)$, and therefore its contribution to the probability is time independent.
	Since $\Tr\mathcal{N}^{(ar)}=\delta_{ar}\Tr\rho_{\beta}^{1/2}$, using \eqref{eqB13}, the contribution of the identity part is exact:
	\begin{equation}
		\overline{P}_{I,\beta}=\dfrac{1}{d_{A}^{3}d_{D}}\sum_{a,r}\overline{\left\|\Tr_{D}\mathcal{N}^{(ar)}_{I}\right\|_{F}^{2}}=\dfrac{1}{d_A^2d_B}\overline{\left|\Tr\rho_{\beta}^{1/2}\right|^{2}}.
	\end{equation}
	It is therefore natural to introduce the thermal quantity
	\begin{equation}
		\label{eq:eta_xi_AD}
		\eta_{\beta}:=\dfrac{1}{d_A^2d_B}\overline{\left|\Tr\rho_{\beta}^{1/2}\right|^{2}}.
	\end{equation}
	With this definition, $\overline{P}_{I,\beta}=\eta_{\beta}$.
	
	We notice that the cross term between $\mathcal{N}^{(ar)}_{I}$ and $\mathcal{N}^{(ar)}_{0}$ in \eqref{eqB12} vanishes exactly, because the cross term is proportional to
	\begin{equation}
		\Tr_{C}[\Tr_{D}\mathcal{N}_{0}^{(ar)}(t)]=\Tr(U\mathcal{N}_{0}^{(ar)}U^\dagger)=0.
	\end{equation}
	Hence, the postselection probability simply decomposes into the identity and traceless parts, such that
	\begin{equation}
		\overline{P_{\beta}(t)} = \overline{P_{I, \beta}(t)} + \overline{P_{0, \beta}(t)},
	\end{equation}
	so we have
	\begin{equation}
		\label{eqB19}
		\overline{P_{0,\beta}(t)}=\dfrac{1}{d_{A}^{3}d_{D}}\sum_{a,r}\overline{\left\|\Tr_{D}\mathcal{N}^{(ar)}_{0}(t)\right\|_{F}^{2}}.
	\end{equation}
	
	At late times after scrambling, we estimate this traceless contribution using the operator expansion of $\mathcal{N}^{(ar)}_{0}(t)$.
	We choose an orthonormal operator basis on $B=CD$, denoted by $\{T_{\mu}\}_{\mu=0}^{d_B^2-1}$, where $T_0=I_B/\sqrt{d_B}$ and $\Tr(T_{\mu}^{\dagger}T_{\nu})=\delta_{\mu\nu}$.
	We then expand $\mathcal{N}^{(ar)}_{0}(t)$ in the orthonormal basis,
	\begin{equation}
		\label{eqB20p}
		\mathcal{N}_{0}^{(ar)}(t)=\sum_{\mu=1}^{d_B^2-1}c_{\mu}^{(ar)}(t)T_{\mu}.
	\end{equation}
	Since this is traceless, we have $c_0^{(ar)}(t)=0$.
	To rewrite \eqref{eqB19} in terms of the expansion coefficients, we rewrite $T_{\mu}=T_{\nu}^{(C)}\otimes T_{\alpha}^{(D)}$, where $\{T_{\nu}^{(C)}\}_{\nu=0}^{d_C^2-1}$ and $\{T_{\alpha}^{(D)}\}_{\alpha=0}^{d_D^2-1}$ are orthonormal operator bases with $T_0^{(C)}=I_C/\sqrt{d_C}$ and $T_0^{(D)}=I_D/\sqrt{d_D}$.
	Then the expansion can be written as
	\begin{equation}
		\mathcal{N}_{0}^{(ar)}(t)=\sum_{(\nu, \alpha)\neq(0,0)}c_{\nu\alpha}^{(ar)}(t)T_{\nu}^{(C)}\otimes T_{\alpha}^{(D)}.
	\end{equation}
	Taking the partial trace over $D$, all terms with $\alpha\neq0$ vanish, while $\Tr_D T_0^{(D)}=\sqrt{d_D}$.
	Therefore,
	\begin{equation}
		\Tr_D\mathcal{N}_{0}^{(ar)}(t)=\sqrt{d_D}\sum_{\nu=1}^{d_C^2-1}c_{\nu0}^{(ar)}(t)T_{\nu}^{(C)}.
	\end{equation}
	Using the orthonormality of $T_{\nu}^{(C)}$, we obtain
	\begin{equation}
		\left\|\Tr_D\mathcal{N}_{0}^{(ar)}(t)\right\|_F^2=d_D\sum_{\nu=1}^{d_C^2-1}|c_{\nu0}^{(ar)}(t)|^2.
	\end{equation}
	Substituting this into \eqref{eqB19}, we find
	\begin{equation}
		\overline{P_{0,\beta}(t)}=\dfrac{1}{d_A^3}\sum_{a,r}\sum_{\nu=1}^{d_C^2-1}\overline{|c_{\nu0}^{(ar)}(t)|^2}.
	\end{equation}
	Equivalently, if $S_C$ denotes the surviving traceless directions $T_{\nu}^{(C)}\otimes T_{0}^{(D)}$ with $\nu=1,\ldots,d_C^2-1$, this becomes
	\begin{equation}
		\label{eqB25}
		\overline{P_{0,\beta}(t)}=\dfrac{1}{d_A^3}\sum_{a,r}\sum_{\mu\in S_C}\overline{|c_{\mu}^{(ar)}(t)|^2}.
	\end{equation}
	
	To evaluate this, we work within a fixed symmetry sector and assume that there is no residual degeneracy within that sector.
	In the scrambled regime, we then assume that the disorder-averaged coefficient weights are uniform across the traceless operator-basis directions, namely, from \eqref{eqB20p},
		\begin{equation}
			\label{eqB21}
			\overline{|c_{\mu}^{(ar)}(t)|^2}\approx\dfrac{\overline{\|\mathcal{N}_{0}^{(ar)}\|_{F}^{2}}}{d_B^2-1}.
		\end{equation}
		This means that, after scrambling and disorder averaging, the traceless operator has no preferred basis direction in the traceless operator space.\footnote{A related result for Haar-random unitaries is discussed in Ref.~\cite{PhysRevA.107.032418}.}
		The underlying assumption is the uniform spreading of traceless operators in the scrambled regime, which may also hold approximately for a single sufficiently chaotic Hamiltonian with large system size.
		Since the number of surviving directions in $S_C$ is $d_C^2-1$, applying \eqref{eqB21} to \eqref{eqB25} gives
		\begin{equation}
			\label{eqB27}
			\overline{P}_{0,\beta}\approx\dfrac{d_C^2-1}{d_{A}^{3}(d_B^2-1)}\sum_{a,r}\overline{\|\mathcal{N}^{(ar)}_0\|_F^2}.
		\end{equation}
		
		To compute the right-hand side, it is convenient to compute the quantity $\|\mathcal{N}^{(ar)}\|_{F}^{2}$, rather than directly solving its traceless part.
		This can be evaluated exactly.
		Using \eqref{eq:C_Nexplicit}, we find $\|\mathcal{N}^{(ar)}\|_{F}^{2} = \|\Tr_{b}\rho_{\beta}^{1/2}\|_{F}^{2}$.
		Hence,
		\begin{equation}
			\label{eqB20}
			\sum_{a,r}\overline{\|\mathcal{N}^{(ar)}\|_{F}^{2}} = d_A^3\xi_{\beta},
		\end{equation}
		where we have defined
		\begin{equation}
			\xi_{\beta}:=\dfrac{1}{d_A}\overline{\left\|\Tr_b\rho_{\beta}^{1/2}\right\|_{F}^{2}}.
		\end{equation}
		Now we consider the identity contribution in \eqref{eqB20}.
		Because $\Tr\mathcal{N}^{(ar)}=\delta_{ar}\Tr\rho_{\beta}^{1/2}$, this is
		\begin{equation}
			\label{eqB22}
			\sum_{a,r}\overline{\|\mathcal{N}_{I}^{(ar)}\|_{F}^{2}} = \sum_{a,r}\overline{\dfrac{|\Tr\mathcal{N}^{(ar)}|^{2}}{d_B}}=d_A^3\eta_{\beta}.
		\end{equation}
		Therefore, since the cross term does not contribute, the traceless contribution can be obtained by subtracting the identity contribution \eqref{eqB22} from the total norm \eqref{eqB20}, which gives
		\begin{equation}
			\sum_{a,r}\overline{\|\mathcal{N}^{(ar)}_{0}\|_{F}^{2}}=d_A^3\left(\xi_{\beta}-\eta_{\beta}\right).
		\end{equation}
		Substituting this into \eqref{eqB27}, we find that the contribution of the traceless part is
		\begin{equation}
			\overline{P}_{0,\beta}\approx\left(\xi_{\beta}-\eta_{\beta}\right)\kappa,\qquad \kappa:=\dfrac{d_C^2-1}{d_B^2-1}.
		\end{equation}
		Combining this and the identity sector \eqref{eq:eta_xi_AD} gives
		\begin{equation}
			\label{eq:P_sat_beta_AD}
			\overline{P}_{\beta}\approx\eta_{\beta}+\left(\xi_{\beta}-\eta_{\beta}\right)\kappa.
		\end{equation}
		The temperature dependence is encoded in the two thermal quantities $\eta_{\beta}$ and $\xi_{\beta}$, which are determined by the thermal weight $\rho_{\beta}^{1/2}$ and its reduction over the injected mode.
		
		In the high-temperature limit ($\beta\ll1$), we have $\rho_{\beta}^{1/2}\approx I_{B}/\sqrt{d_{B}}$, so we find
		\begin{equation}
			\eta_{\beta\to0}=\dfrac{1}{d_A^2},\qquad \xi_{\beta\to0}=1.
		\end{equation}
		Thus, \eqref{eq:P_sat_beta_AD} becomes
		\begin{equation}
			\label{eq:B29}
			\overline{P}_{\beta\to0}\approx\dfrac{1}{d_A^2}+\dfrac{d_A^2-1}{d_A^2}\dfrac{d_C^2-1}{d_B^2-1}.
		\end{equation}
		Since $d_B=d_Dd_C$, the correction is $O(1/d_D^2)$.
		Thus, for large $d_D$, the postselection probability approaches $1/d_A^2$, which reproduces the conventional YK protocol at high temperature.
		
		The high-temperature limit can also be seen directly at the operator level.
		At $\beta\ll1$, we see $\mathcal{N}^{(ar)}\to\mathcal{F}^{(ar)}/\sqrt{d_B}$ where
		\begin{equation}
			\mathcal{F}^{(ar)}:=\sum_{n=0}^{d_A-1}\mathcal{W}_{b}^{(na)}\mathcal{W}_{b}^{(nr)\dagger}=d_A\outerproduct{a}{r}_{b}\otimes I_{\bar B},
		\end{equation}
		which satisfies $\Tr\mathcal{F}^{(ar)}=d_B\delta_{ar}$ and $\|\mathcal{F}^{(ar)}\|_{F}^{2}=d_A d_B$.
		Thus, the identity part is
		\begin{equation}
			\mathcal{F}_{I}^{(ar)}=\delta_{ar}I_{B},
		\end{equation}
		while the traceless part satisfies
		\begin{equation}
			\sum_{a,r}\|\mathcal{F}_{0}^{(ar)}\|_{F}^{2}=d_A d_B(d_A^2-1).
		\end{equation}
		Substituting these identities into the finite-temperature formula gives \eqref{eq:B29}.
		
		The initial value at $t=0$ can also be obtained directly from the same block representation.
		At $t=0$, we have $U=I$.
		Using \eqref{eq:C_Nexplicit},
		\begin{equation}
			\Tr_{D}\mathcal{N}^{(ar)}=\outerproduct{a}{r}_{b}\otimes\Tr_{D}\Tr_{b}\rho_{\beta}^{1/2}.
		\end{equation}
		If the injected qubit belongs to $C$, then
		\begin{equation}
			\label{eq:C27}
			P_{\beta}(0)=\dfrac{1}{d_A d_D}\left\|\Tr_{bD}\rho_{\beta}^{1/2}\right\|_{F}^{2}.
		\end{equation}
		In the high-temperature limit, this becomes
		\begin{equation}
			P_{\beta\to0}(0)=1.
		\end{equation}
		Thus, the postselection probability starts from unity in the limit $\beta\to0$ and then decreases toward the late-time saturation value after scrambling.
		
		\section{\label{appendix:C}Conditional fidelity at finite temperature}
		We now evaluate the asymptotic value of the conditional fidelity.
		The calculation parallels Appendix~\ref{appendix:B}.
		The conditional fidelity is defined by
		\begin{equation}
			\label{eqC1}
			F_{\beta}(t)=\dfrac{Q_{\beta}(t)}{P_{\beta}(t)}.
		\end{equation}
		Here $P_{\beta}(t)$ is the postselection probability derived in the previous section, while $Q_{\beta}(t)$ additionally projects the reference systems $RR'$ onto the EPR state:
		\begin{equation}
			Q_{\beta}(t)=\langle\widetilde{\text{out}}(t)|\Pi_{AA'}\Pi_{DD'}\Pi_{RR'}|\widetilde{\text{out}}(t)\rangle.
		\end{equation}
		Using \eqref{eqB4}, $\Pi_{AA'}$ and $\Pi_{RR'}$ give rise to the factors $\delta_{a'r'}$ and $\delta_{ar}$, respectively.
		Then, following the same procedures \eqref{eqB5}--\eqref{eqB9}, we obtain
		\begin{equation}
			Q_{\beta}(t) = \dfrac{1}{d_{A}^{4}d_{D}^{2}}\sum_{\alpha}\Tr[\mathcal{P}_{\alpha}\mathcal{S}(t)\mathcal{P}_{\alpha}\mathcal{S}(t)^{\dagger}],
		\end{equation}
		where we have defined
		\begin{equation}
			\mathcal{S}:=\sum_{m,n=0}^{d_A-1}\mathcal{W}_{b}^{(mn)}\rho_{\beta}^{1/2}\mathcal{W}_{b}^{(mn)\dagger},\quad \mathcal{S}(t)=U(t)\mathcal{S}U^{\dagger}(t),
		\end{equation}
		analogous to \eqref{eq:B_Nblock}.
		Using \eqref{eqB10} and taking a disorder average, we find
		\begin{equation}
			\label{eqC5}
			\overline{Q_{\beta}(t)}=\dfrac{1}{d_A^4d_D}\overline{\left\|\Tr_D\mathcal{S}(t)\right\|_F^2}.
		\end{equation}
		To compute this, we decompose $\mathcal{S}=\mathcal{S}_{I}+\mathcal{S}_{0}$ with
		\begin{equation}
			\mathcal{S}_{I}=\dfrac{\Tr\mathcal{S}}{d_B}I_B,\qquad \Tr\mathcal{S}_{0}=0.
		\end{equation}
		The identity part $\mathcal{S}_{I}\propto I_B$ is invariant under $U(t)$, and the cross term between $\mathcal{S}_{I}$ and $\mathcal{S}_{0}$ in \eqref{eqC5} vanishes for the same reason as before.
		Then, we can also decompose
		\begin{equation}
			\overline{Q_{\beta}(t)} = \overline{Q_{I, \beta}(t)} + \overline{Q_{0, \beta}(t)},
		\end{equation}
		where each term represents identity and traceless part, respectively.
		We first focus on the identity part.
		Using \eqref{eq:C5}, one obtains the relation $\mathcal{S}=I_{b}\otimes\Tr_{b}\rho_{\beta}^{1/2}$, such that $\Tr\mathcal{S}=d_A\Tr\rho_{\beta}^{1/2}$.
		Therefore, the identity contribution is exact:
		\begin{equation}
			\label{eqC7}
			\overline{Q}_{I,\beta}=\dfrac{1}{d_A^4d_D}\overline{\left\|\Tr_D\mathcal{S}_{I}\right\|_F^2}=\dfrac{1}{d_A^2d_B}\overline{\left|\Tr\rho_{\beta}^{1/2}\right|^2}=\eta_{\beta}.
		\end{equation}
		
		For the traceless contribution from $\mathcal{S}_{0}$, we subtract the identity part from the total norm.
		The total norm is
		\begin{equation}
			\overline{\left\|\mathcal{S}\right\|_F^2}=d_A^2\xi_{\beta},
		\end{equation}
		while the identity part gives
		\begin{equation}
			\overline{\left\|\mathcal{S}_{I}\right\|_F^2}=d_A^4\eta_{\beta}.
		\end{equation}
		Therefore,
		\begin{equation}
			\overline{\left\|\mathcal{S}_{0}\right\|_F^2}=\overline{\left\|\mathcal{S}\right\|_F^2}-\overline{\left\|\mathcal{S}_{I}\right\|_F^2}=d_A^4\left(\dfrac{\xi_{\beta}}{d_A^2}-\eta_{\beta}\right).
		\end{equation}
		Following \eqref{eqB20p}--\eqref{eqB27}, assuming the same scrambled second-moment estimate as in \eqref{eqB21}, this results in
		\begin{equation}
			\overline{Q}_{0,\beta}\approx\dfrac{\kappa}{d_A^4}\overline{\left\|\mathcal{S}_{0}\right\|_F^2}=\left(\dfrac{\xi_{\beta}}{d_A^2}-\eta_{\beta}\right)\kappa.
		\end{equation}
		Combining this and \eqref{eqC7}, we obtain
		\begin{equation}
			\label{eq:Q_sat_beta_AD}
			\overline{Q}_{\beta}\approx\eta_{\beta}+\left(\dfrac{\xi_{\beta}}{d_A^2}-\eta_{\beta}\right)\kappa.
		\end{equation}
		Therefore, using \eqref{eqC1} and assuming that the late-time sample-to-sample fluctuations of $P_{\beta}(t)$ and $Q_{\beta}(t)$ are sufficiently small, the disorder-averaged conditional fidelity is estimated as
		\begin{equation}
			\label{eq:F_sat_beta_AD_ratio}
			\overline{F}_{\beta}\approx\dfrac{\eta_{\beta}+\left(\dfrac{\xi_{\beta}}{d_A^2}-\eta_{\beta}\right)\kappa}{\eta_{\beta}+\left(\xi_{\beta}-\eta_{\beta}\right)\kappa}.
		\end{equation}
		Equivalently, using \eqref{eq:P_sat_beta_AD},
		\begin{equation}
			\label{eq:F_sat_beta_AD}
			\overline{F}_{\beta}\approx1 - \kappa\left(1-\dfrac{1}{d_A^2}\right)\dfrac{\xi_{\beta}}{\overline{P}_{\beta}}.
		\end{equation}
		
		In the high-temperature limit, $\eta_{\beta}\to1/d_A^2$ and $\xi_{\beta}\to1$.
		Thus,
		\begin{equation}
			\overline{Q}_{\beta\to0}\approx\dfrac{1}{d_A^2},
		\end{equation}
		and the fidelity reduces to
		\begin{equation}
			\overline{F}_{\beta\to0}\approx\dfrac{1}{d_A^2\overline{P}_{\beta\to0}}\approx\dfrac{1}{1+(d_A^2-1)\kappa}.
		\end{equation}
		Therefore, the high-$T$ fidelity approaches unity when $N_{D}$ is large.
		
		We also compute the initial value of the fidelity.
		At $t=0$, assuming that the injected qubit on $b$ belongs to $C$, we have
		\begin{equation}
			Q_{\beta}(0)=\dfrac{1}{d_A^3d_D}\left\|\Tr_{bD}\rho_{\beta}^{1/2}\right\|_{F}^{2}.
		\end{equation}
		Using the initial postselection probability \eqref{eq:C27}, we obtain
		\begin{equation}
			F_{\beta}(0)=\dfrac{Q_{\beta}(0)}{P_{\beta}(0)}=\dfrac{1}{d_A^2},
		\end{equation}
		which is independent of $T$.
		For the one-qubit diary used in this work, this gives $F_{\beta}(0)=1/4$.
		Thus, including the expelled qubit in the radiation does not change the trivial initial EPR overlap; the conditional fidelity increases only after the scrambling dynamics transfers the injected diary information into the radiation degrees of freedom accessible to the decoder.
		

		\section{\label{appendix:isotropic_diagnostic}Numerical diagnostic of uniform spreading}
		In Appendices~\ref{appendix:B} and \ref{appendix:C}, the late-time saturation values rely on the uniform-spreading assumption in \eqref{eqB21}.
		We numerically examine this assumption by visualizing the coefficients $c_{\mu}^{(ar)}(t)$ defined in \eqref{eqB20p}.
		For a fixed block $(a,r)$, we define the normalized density
		\begin{equation}
			p_{\nu\alpha}^{(ar)}(t) := \dfrac{\overline{|c_{\nu\alpha}^{(ar)}(t)|^2}}{\overline{\|\mathcal{N}_{0}^{(ar)}\|_{F}^{2}}}.
		\end{equation}
		The uniform-spreading assumption in \eqref{eqB21} then predicts
		\begin{equation}
			\label{eqD2}
			g_{\nu}^{(ar)}(t)
			:=
			(d_B^2-1)p_{\nu0}^{(ar)}(t)-1
			\approx 0
		\end{equation}
		at late times for the traceless directions with $\alpha=0$ and $\nu\neq0$.
		
		\begin{figure}[tb]
			\centering
			\includegraphics[width=\linewidth]{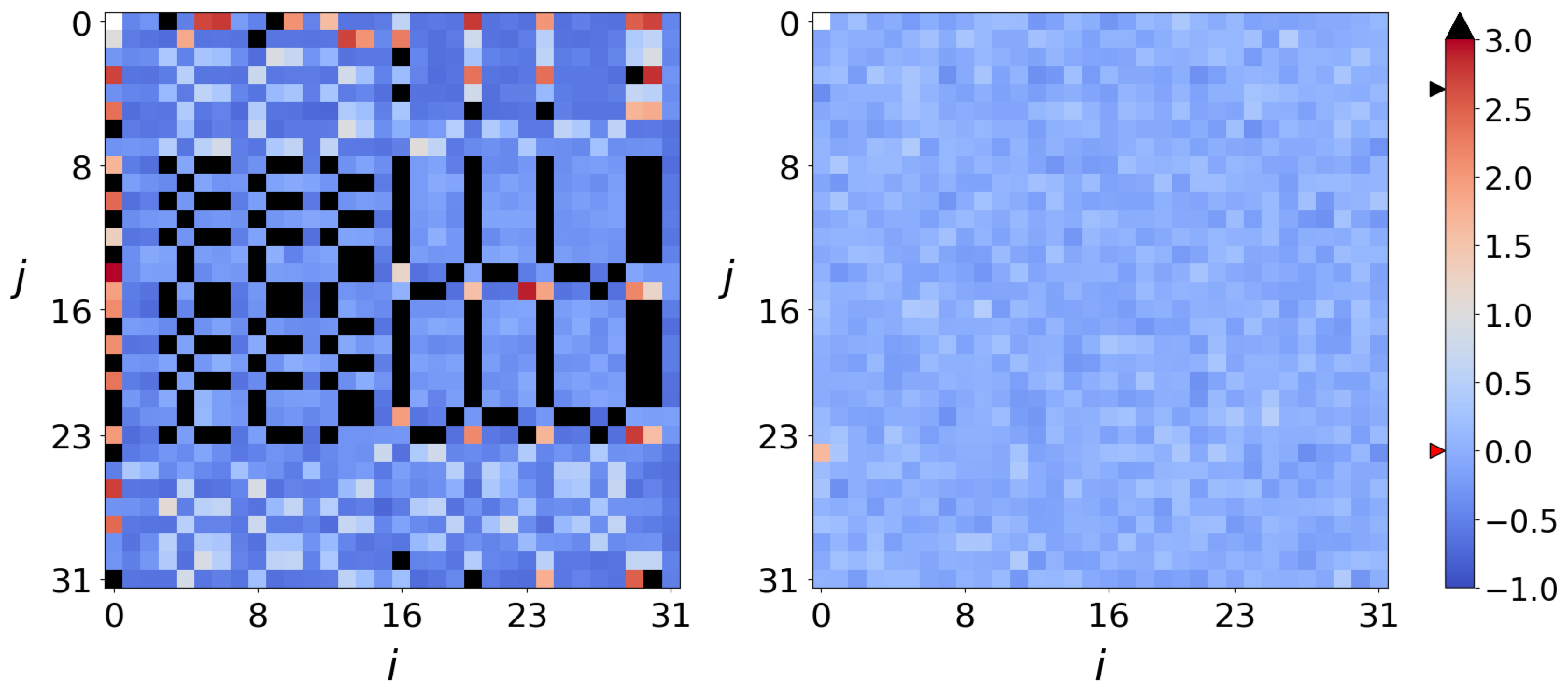}
			\caption{
				\label{fig:isotropic_spreading}
				Numerical diagnostic of the uniform-spreading assumption at $T=1$ for  $t=2$ (left) and $t=10$ (right).
				The panels show $g_{\nu}^{(00)}(t)$ over 100 disorder realizations for the $N=16$ SYK model with $N_A=1$, $N_B=7$, $N_C=5$, and $N_D=2$.  
				The index $\nu$ is displayed as a two-dimensional array via $\nu(i,j)=i d_C+j$.
				Values near zero correspond to the uniform-spreading prediction, while black indicates values above the displayed range.
				Triangles to the left of the color bar denote the mean plotted value at early time (black) and late time (red).
			}
		\end{figure}
		
		In Fig.~\ref{fig:isotropic_spreading}, we show $g_{\nu}^{(00)}(t)$ for the $N=16$ SYK model with $N_A=1$, $N_B=7$, $N_C=5$, and $N_D=2$, averaged over 100 disorder realizations.
		We compare the early- and late-time results at $t=2$ and $t=10$, respectively, with the temperature fixed at $T = 1$.
		To represent the one-dimensional index $\nu$ in matrix form, we reshape $g_{\nu}^{(00)}(t)$ into matrix elements $g_{ij}^{(00)}(t)$, where the row and column indices $(i,j)$ are related to $\nu$ by $\nu(i,j)=i d_C+j$.
		In the present case, $d_C^2=1024$, and hence the coefficients are reshaped into a $32\times32$ matrix corresponding to $\nu=0,\ldots,1023$.
		The entry $\nu=0$, which represents the identity direction, is excluded from the traceless component.
		
		At late times, the matrix elements cluster around zero, whereas substantial deviations remain at early times.
		This behavior supports \eqref{eqB21} as an approximate late-time description of the $\alpha=0$ traceless sector under sufficiently scrambling dynamics.
		
		
		\bibliographystyle{jhep}
		\bibliography{ref}

@article{Hayden2007,
	doi = {10.1088/1126-6708/2007/09/120},
	url = {https://doi.org/10.1088/1126-6708/2007/09/120},
	year = {2007},
	month = {sep},
	publisher = {},
	volume = {2007},
	number = {09},
	pages = {120},
	author = {Patrick Hayden and John Preskill},
	title = {Black holes as mirrors: quantum information in random subsystems},
	journal = {Journal of High Energy Physics}
}

@article{PhysRevD.106.046011,
	title = {Hayden-Preskill protocol and decoding Hawking radiation at finite temperature},
	author = {Li, Ran and Wang, Jin},
	journal = {Phys. Rev. D},
	volume = {106},
	issue = {4},
	pages = {046011},
	numpages = {17},
	year = {2022},
	month = {Aug},
	publisher = {American Physical Society},
	doi = {10.1103/PhysRevD.106.046011},
	url = {https://link.aps.org/doi/10.1103/PhysRevD.106.046011}
}

@misc{yoshida2017efficientdecodinghaydenpreskillprotocol,
	title={Efficient decoding for the Hayden-Preskill protocol}, 
	author={Beni Yoshida and Alexei Kitaev},
	year={2017},
	eprint={1710.03363},
	archivePrefix={arXiv},
	primaryClass={hep-th},
	url={https://arxiv.org/abs/1710.03363}, 
}

@article{Caceres2021,
	author = {C{\'a}ceres, Elena and Misobuchi, Anderson and Pimentel, Rafael},
	date = {2021/11/04},
	doi = {10.1007/JHEP11(2021)015},
	id = {C{\'a}ceres2021},
	isbn = {1029-8479},
	journal = {Journal of High Energy Physics},
	number = {11},
	pages = {15},
	title = {Sparse SYK and traversable wormholes},
	url = {https://doi.org/10.1007/JHEP11(2021)015},
	volume = {2021},
	year = {2021}
}

@article{Gao2017,
	author = {Gao, Ping and Jafferis, Daniel Louis and Wall, Aron C.},
	date = {2017/12/28},
	doi = {10.1007/JHEP12(2017)151},
	id = {Gao2017},
	isbn = {1029-8479},
	journal = {Journal of High Energy Physics},
	number = {12},
	pages = {151},
	title = {Traversable wormholes via a double trace deformation},
	url = {https://doi.org/10.1007/JHEP12(2017)151},
	volume = {2017},
	year = {2017}
}

@article{PhysRevA.104.012427,
	title = {Variational preparation of the thermofield double state of the Sachdev-Ye-Kitaev model},
	author = {Su, Vincent Paul},
	journal = {Phys. Rev. A},
	volume = {104},
	issue = {1},
	pages = {012427},
	numpages = {13},
	year = {2021},
	month = {Jul},
	publisher = {American Physical Society},
	doi = {10.1103/PhysRevA.104.012427},
	url = {https://link.aps.org/doi/10.1103/PhysRevA.104.012427}
}

@article{Jafferis2022,
	author = {Jafferis, Daniel and Zlokapa, Alexander and Lykken, Joseph D. and Kolchmeyer, David K. and Davis, Samantha I. and Lauk, Nikolai and Neven, Hartmut and Spiropulu, Maria},
	date = {2022/12/01},
	doi = {10.1038/s41586-022-05424-3},
	id = {Jafferis2022},
	isbn = {1476-4687},
	journal = {Nature},
	number = {7938},
	pages = {51--55},
	title = {Traversable wormhole dynamics on a quantum processor},
	url = {https://doi.org/10.1038/s41586-022-05424-3},
	volume = {612},
	year = {2022}}

@article{PRXQuantum.4.010320,
	title = {Quantum Gravity in the Lab. I. Teleportation by Size and Traversable Wormholes},
	author = {Brown, Adam R. and Gharibyan, Hrant and Leichenauer, Stefan and Lin, Henry W. and Nezami, Sepehr and Salton, Grant and Susskind, Leonard and Swingle, Brian and Walter, Michael},
	journal = {PRX Quantum},
	volume = {4},
	issue = {1},
	pages = {010320},
	numpages = {22},
	year = {2023},
	month = {Feb},
	publisher = {American Physical Society},
	doi = {10.1103/PRXQuantum.4.010320},
	url = {https://link.aps.org/doi/10.1103/PRXQuantum.4.010320}
}

@article{PhysRevLett.70.3339,
	title = {Gapless spin-fluid ground state in a random quantum Heisenberg magnet},
	author = {Sachdev, Subir and Ye, Jinwu},
	journal = {Phys. Rev. Lett.},
	volume = {70},
	issue = {21},
	pages = {3339--3342},
	numpages = {0},
	year = {1993},
	month = {May},
	publisher = {American Physical Society},
	doi = {10.1103/PhysRevLett.70.3339},
	url = {https://link.aps.org/doi/10.1103/PhysRevLett.70.3339}
}

@article{Maldacena2016,
	author = {Maldacena, Juan and Shenker, Stephen H. and Stanford, Douglas},
	date = {2016/08/17},
	doi = {10.1007/JHEP08(2016)106},
	id = {Maldacena2016},
	isbn = {1029-8479},
	journal = {Journal of High Energy Physics},
	number = {8},
	pages = {106},
	title = {A bound on chaos},
	url = {https://doi.org/10.1007/JHEP08(2016)106},
	volume = {2016},
	year = {2016}}

@article{PhysRevLett.117.111601,
	title = {Chaos in ${\mathrm{AdS}}_{2}$ Holography},
	author = {Jensen, Kristan},
	journal = {Phys. Rev. Lett.},
	volume = {117},
	issue = {11},
	pages = {111601},
	numpages = {6},
	year = {2016},
	month = {Sep},
	publisher = {American Physical Society},
	doi = {10.1103/PhysRevLett.117.111601},
	url = {https://link.aps.org/doi/10.1103/PhysRevLett.117.111601}
}

@misc{xu2020sparsemodelquantumholography,
	title={A Sparse Model of Quantum Holography}, 
	author={Shenglong Xu and Leonard Susskind and Yuan Su and Brian Swingle},
	year={2020},
	eprint={2008.02303},
	archivePrefix={arXiv},
	primaryClass={cond-mat.str-el},
	url={https://arxiv.org/abs/2008.02303}, 
}

@article{PhysRevD.103.106002,
	title = {Sparse Sachdev-Ye-Kitaev model, quantum chaos, and gravity duals},
	author = {Garc\'{\i}a-Garc\'{\i}a, Antonio M. and Jia, Yiyang and Rosa, Dario and Verbaarschot, Jacobus J. M.},
	journal = {Phys. Rev. D},
	volume = {103},
	issue = {10},
	pages = {106002},
	numpages = {28},
	year = {2021},
	month = {May},
	publisher = {American Physical Society},
	doi = {10.1103/PhysRevD.103.106002},
	url = {https://link.aps.org/doi/10.1103/PhysRevD.103.106002}
}

@article{PhysRevB.107.L081103,
	title = {Binary-coupling sparse Sachdev-Ye-Kitaev model: An improved model of quantum chaos and holography},
	author = {Tezuka, Masaki and Oktay, Onur and Rinaldi, Enrico and Hanada, Masanori and Nori, Franco},
	journal = {Phys. Rev. B},
	volume = {107},
	issue = {8},
	pages = {L081103},
	numpages = {7},
	year = {2023},
	month = {Feb},
	publisher = {American Physical Society},
	doi = {10.1103/PhysRevB.107.L081103},
	url = {https://link.aps.org/doi/10.1103/PhysRevB.107.L081103}
}

@article{Orman2025,
	author = {Orman, Patrick and Gharibyan, Hrant and Preskill, John},
	date = {2025/02/26},
	doi = {10.1007/JHEP02(2025)173},
	id = {Orman2025},
	isbn = {1029-8479},
	journal = {Journal of High Energy Physics},
	number = {2},
	pages = {173},
	title = {Quantum chaos in the sparse SYK model},
	url = {https://doi.org/10.1007/JHEP02(2025)173},
	volume = {2025},
	year = {2025}}

@article{Cotler2017,
	author = {Cotler, Jordan and Hunter-Jones, Nicholas and Liu, Junyu and Yoshida, Beni},
	date = {2017/11/09},
	doi = {10.1007/JHEP11(2017)048},
	id = {Cotler2017},
	isbn = {1029-8479},
	journal = {Journal of High Energy Physics},
	number = {11},
	pages = {48},
	title = {Chaos, complexity, and random matrices},
	url = {https://doi.org/10.1007/JHEP11(2017)048},
	volume = {2017},
	year = {2017}}

@article{PhysRevLett.110.084101,
	title = {Distribution of the Ratio of Consecutive Level Spacings in Random Matrix Ensembles},
	author = {Atas, Y. Y. and Bogomolny, E. and Giraud, O. and Roux, G.},
	journal = {Phys. Rev. Lett.},
	volume = {110},
	issue = {8},
	pages = {084101},
	numpages = {5},
	year = {2013},
	month = {Feb},
	publisher = {American Physical Society},
	doi = {10.1103/PhysRevLett.110.084101},
	url = {https://link.aps.org/doi/10.1103/PhysRevLett.110.084101}
}

@article{PRXQuantum.4.010321,
	title = {Quantum Gravity in the Lab. II. Teleportation by Size and Traversable Wormholes},
	author = {Nezami, Sepehr and Lin, Henry W. and Brown, Adam R. and Gharibyan, Hrant and Leichenauer, Stefan and Salton, Grant and Susskind, Leonard and Swingle, Brian and Walter, Michael},
	journal = {PRX Quantum},
	volume = {4},
	issue = {1},
	pages = {010321},
	numpages = {33},
	year = {2023},
	month = {Feb},
	publisher = {American Physical Society},
	doi = {10.1103/PRXQuantum.4.010321},
	url = {https://link.aps.org/doi/10.1103/PRXQuantum.4.010321}
}

@article{82edc856-4d85-3b98-9b0d-ad55bb9315f6,
	ISSN = {00029939, 10886826},
	URL = {http://www.jstor.org/stable/2033649},
	author = {H. F. Trotter},
	journal = {Proceedings of the American Mathematical Society},
	number = {4},
	pages = {545--551},
	publisher = {American Mathematical Society},
	title = {On the Product of Semi-Groups of Operators},
	urldate = {2026-03-29},
	volume = {10},
	year = {1959}
}

@article{Kandala2017,
	author = {Kandala, Abhinav and Mezzacapo, Antonio and Temme, Kristan and Takita, Maika and Brink, Markus and Chow, Jerry M. and Gambetta, Jay M.},
	date = {2017/09/01},
	doi = {10.1038/nature23879},
	id = {Kandala2017},
	isbn = {1476-4687},
	journal = {Nature},
	number = {7671},
	pages = {242--246},
	title = {Hardware-efficient variational quantum eigensolver for small molecules and quantum magnets},
	url = {https://doi.org/10.1038/nature23879},
	volume = {549},
	year = {2017}}

@article{
	Seth1996,
	author = {Seth Lloyd },
	title = {Universal Quantum Simulators},
	journal = {Science},
	volume = {273},
	number = {5278},
	pages = {1073-1078},
	year = {1996},
	doi = {10.1126/science.273.5278.1073},
	URL = {https://www.science.org/doi/abs/10.1126/science.273.5278.1073}
}

@article{PhysRevX.12.031013,
	title = {Many-Body Quantum Teleportation via Operator Spreading in the Traversable Wormhole Protocol},
	author = {Schuster, Thomas and Kobrin, Bryce and Gao, Ping and Cong, Iris and Khabiboulline, Emil T. and Linke, Norbert M. and Lukin, Mikhail D. and Monroe, Christopher and Yoshida, Beni and Yao, Norman Y.},
	journal = {Phys. Rev. X},
	volume = {12},
	issue = {3},
	pages = {031013},
	numpages = {61},
	year = {2022},
	month = {Jul},
	publisher = {American Physical Society},
	doi = {10.1103/PhysRevX.12.031013},
	url = {https://link.aps.org/doi/10.1103/PhysRevX.12.031013}
}

@article{Gharibyan2018,
	author = {Gharibyan, Hrant and Hanada, Masanori and Shenker, Stephen H. and Tezuka, Masaki},
	date = {2018/07/18},
	doi = {10.1007/JHEP07(2018)124},
	id = {Gharibyan2018},
	isbn = {1029-8479},
	journal = {Journal of High Energy Physics},
	number = {7},
	pages = {124},
	title = {Onset of random matrix behavior in scrambling systems},
	url = {https://doi.org/10.1007/JHEP07(2018)124},
	volume = {2018},
	year = {2018}}

@article{PhysRevD.94.106002,
	title = {Remarks on the Sachdev-Ye-Kitaev model},
	author = {Maldacena, Juan and Stanford, Douglas},
	journal = {Phys. Rev. D},
	volume = {94},
	issue = {10},
	pages = {106002},
	numpages = {43},
	year = {2016},
	month = {Nov},
	publisher = {American Physical Society},
	doi = {10.1103/PhysRevD.94.106002},
	url = {https://link.aps.org/doi/10.1103/PhysRevD.94.106002}
}

@misc{Kitaev2015,
	author       = {Kitaev, Alexei},
	title        = {A Simple Model of Quantum Holography},
	howpublished = {KITP Strings Seminar and Entanglement 2015 Program},
	year         = {2015},
	note         = {Talks given on February 12, April 7, and May 27, 2015},
	url          = {http://online.kitp.ucsb.edu/online/entangled15/}
}

@article{Gao2021,
	author = {Gao, Ping and Jafferis, Daniel Louis},
	date = {2021/07/15},
	doi = {10.1007/JHEP07(2021)097},
	id = {Gao2021},
	isbn = {1029-8479},
	journal = {Journal of High Energy Physics},
	number = {7},
	pages = {97},
	title = {A traversable wormhole teleportation protocol in the SYK model},
	url = {https://doi.org/10.1007/JHEP07(2021)097},
	volume = {2021},
	year = {2021}}

@article{PhysRevD.98.086026,
	title = {Spectral form factors and late time quantum chaos},
	author = {Liu, Junyu},
	journal = {Phys. Rev. D},
	volume = {98},
	issue = {8},
	pages = {086026},
	numpages = {28},
	year = {2018},
	month = {Oct},
	publisher = {American Physical Society},
	doi = {10.1103/PhysRevD.98.086026},
	url = {https://link.aps.org/doi/10.1103/PhysRevD.98.086026}
}

@article{Granet2026,
	author = {Granet, Etienne and Kikuchi, Yuta and Dreyer, Henrik and Rinaldi, Enrico},
	date = {2026/02/23},
	doi = {10.1038/s41534-026-01206-1},
	id = {Granet2026},
	isbn = {2056-6387},
	journal = {npj Quantum Information},
	number = {1},
	pages = {43},
	title = {Simulating sparse SYK model with a randomized algorithm on a trapped-ion quantum computer},
	url = {https://doi.org/10.1038/s41534-026-01206-1},
	volume = {12},
	year = {2026}}

@article{PhysRevLett.126.030602,
	title = {Many-Body Chaos in the Sachdev-Ye-Kitaev Model},
	author = {Kobrin, Bryce and Yang, Zhenbin and Kahanamoku-Meyer, Gregory D. and Olund, Christopher T. and Moore, Joel E. and Stanford, Douglas and Yao, Norman Y.},
	journal = {Phys. Rev. Lett.},
	volume = {126},
	issue = {3},
	pages = {030602},
	numpages = {6},
	year = {2021},
	month = {Jan},
	publisher = {American Physical Society},
	doi = {10.1103/PhysRevLett.126.030602},
	url = {https://link.aps.org/doi/10.1103/PhysRevLett.126.030602}
}

@article{PhysRevLett.123.220502,
	title = {Variational Thermal Quantum Simulation via Thermofield Double States},
	author = {Wu, Jingxiang and Hsieh, Timothy H.},
	journal = {Phys. Rev. Lett.},
	volume = {123},
	issue = {22},
	pages = {220502},
	numpages = {6},
	year = {2019},
	month = {Nov},
	publisher = {American Physical Society},
	doi = {10.1103/PhysRevLett.123.220502},
	url = {https://link.aps.org/doi/10.1103/PhysRevLett.123.220502}
}

@article{PhysRevX.9.011006,
	title = {Disentangling Scrambling and Decoherence via Quantum Teleportation},
	author = {Yoshida, Beni and Yao, Norman Y.},
	journal = {Phys. Rev. X},
	volume = {9},
	issue = {1},
	pages = {011006},
	numpages = {17},
	year = {2019},
	month = {Jan},
	publisher = {American Physical Society},
	doi = {10.1103/PhysRevX.9.011006},
	url = {https://link.aps.org/doi/10.1103/PhysRevX.9.011006}
}

@article{Roberts2017,
	author = {Roberts, Daniel A. and Yoshida, Beni},
	date = {2017/04/20},
	doi = {10.1007/JHEP04(2017)121},
	id = {Roberts2017},
	isbn = {1029-8479},
	journal = {Journal of High Energy Physics},
	number = {4},
	pages = {121},
	title = {Chaos and complexity by design},
	url = {https://doi.org/10.1007/JHEP04(2017)121},
	volume = {2017},
	year = {2017}}

@article{
	doi:10.1126/science.abg5029,
	author = {Xiao Mi  and Pedram Roushan  and Chris Quintana  and Salvatore Mandrà  and Jeffrey Marshall  and Charles Neill  and Frank Arute  and Kunal Arya  and Juan Atalaya  and Ryan Babbush  and Joseph C. Bardin  and Rami Barends  and Joao Basso  and Andreas Bengtsson  and Sergio Boixo  and Alexandre Bourassa  and Michael Broughton  and Bob B. Buckley  and David A. Buell  and Brian Burkett  and Nicholas Bushnell  and Zijun Chen  and Benjamin Chiaro  and Roberto Collins  and William Courtney  and Sean Demura  and Alan R. Derk  and Andrew Dunsworth  and Daniel Eppens  and Catherine Erickson  and Edward Farhi  and Austin G. Fowler  and Brooks Foxen  and Craig Gidney  and Marissa Giustina  and Jonathan A. Gross  and Matthew P. Harrigan  and Sean D. Harrington  and Jeremy Hilton  and Alan Ho  and Sabrina Hong  and Trent Huang  and William J. Huggins  and L. B. Ioffe  and Sergei V. Isakov  and Evan Jeffrey  and Zhang Jiang  and Cody Jones  and Dvir Kafri  and Julian Kelly  and Seon Kim  and Alexei Kitaev  and Paul V. Klimov  and Alexander N. Korotkov  and Fedor Kostritsa  and David Landhuis  and Pavel Laptev  and Erik Lucero  and Orion Martin  and Jarrod R. McClean  and Trevor McCourt  and Matt McEwen  and Anthony Megrant  and Kevin C. Miao  and Masoud Mohseni  and Shirin Montazeri  and Wojciech Mruczkiewicz  and Josh Mutus  and Ofer Naaman  and Matthew Neeley  and Michael Newman  and Murphy Yuezhen Niu  and Thomas E. O’Brien  and Alex Opremcak  and Eric Ostby  and Balint Pato  and Andre Petukhov  and Nicholas Redd  and Nicholas C. Rubin  and Daniel Sank  and Kevin J. Satzinger  and Vladimir Shvarts  and Doug Strain  and Marco Szalay  and Matthew D. Trevithick  and Benjamin Villalonga  and Theodore White  and Z. Jamie Yao  and Ping Yeh  and Adam Zalcman  and Hartmut Neven  and Igor Aleiner  and Kostyantyn Kechedzhi  and Vadim Smelyanskiy  and Yu Chen },
	title = {Information scrambling in quantum circuits},
	journal = {Science},
	volume = {374},
	number = {6574},
	pages = {1479-1483},
	year = {2021},
	doi = {10.1126/science.abg5029},
	URL = {https://www.science.org/doi/abs/10.1126/science.abg5029}}

@article{PhysRevA.105.032435,
	title = {Quantum advantages for Pauli channel estimation},
	author = {Chen, Senrui and Zhou, Sisi and Seif, Alireza and Jiang, Liang},
	journal = {Phys. Rev. A},
	volume = {105},
	issue = {3},
	pages = {032435},
	numpages = {17},
	year = {2022},
	month = {Mar},
	publisher = {American Physical Society},
	doi = {10.1103/PhysRevA.105.032435},
	url = {https://link.aps.org/doi/10.1103/PhysRevA.105.032435}
}

@article{PhysRevA.80.012304,
	title = {Exact and approximate unitary 2-designs and their application to fidelity estimation},
	author = {Dankert, Christoph and Cleve, Richard and Emerson, Joseph and Livine, Etera},
	journal = {Phys. Rev. A},
	volume = {80},
	issue = {1},
	pages = {012304},
	numpages = {6},
	year = {2009},
	month = {Jul},
	publisher = {American Physical Society},
	doi = {10.1103/PhysRevA.80.012304},
	url = {https://link.aps.org/doi/10.1103/PhysRevA.80.012304}
}

@article{Hosur2016,
	author = {Hosur, Pavan and Qi, Xiao-Liang and Roberts, Daniel A. and Yoshida, Beni},
	date = {2016/02/01},
	doi = {10.1007/JHEP02(2016)004},
	id = {Hosur2016},
	isbn = {1029-8479},
	journal = {Journal of High Energy Physics},
	number = {2},
	pages = {4},
	title = {Chaos in quantum channels},
	url = {https://doi.org/10.1007/JHEP02(2016)004},
	volume = {2016},
	year = {2016}}

@article{Hawking1975,
author = {Hawking, S.  W. },
date = {1975/08/01},
doi = {10.1007/BF02345020},
id = {Hawking1975},
isbn = {1432-0916},
journal = {Communications in Mathematical Physics},
number = {3},
pages = {199--220},
title = {Particle creation by black holes},
url = {https://doi.org/10.1007/BF02345020},
volume = {43},
year = {1975}}

@article{PhysRevLett.71.3743,
	title = {Information in black hole radiation},
	author = {Page, Don N.},
	journal = {Phys. Rev. Lett.},
	volume = {71},
	issue = {23},
	pages = {3743--3746},
	numpages = {0},
	year = {1993},
	month = {Dec},
	publisher = {American Physical Society},
	doi = {10.1103/PhysRevLett.71.3743},
	url = {https://link.aps.org/doi/10.1103/PhysRevLett.71.3743}
}

@article{10.1093/ptep/ptw124,
	author = {Maldacena, Juan and Stanford, Douglas and Yang, Zhenbin},
	title = {Conformal symmetry and its breaking in two-dimensional nearly anti-de Sitter space},
	journal = {Progress of Theoretical and Experimental Physics},
	volume = {2016},
	number = {12},
	pages = {12C104},
	year = {2016},
	month = {12},
	issn = {2050-3911},
	doi = {10.1093/ptep/ptw124},
	url = {https://doi.org/10.1093/ptep/ptw124},
}

@article{PhysRevD.14.2460,
	title = {Breakdown of predictability in gravitational collapse},
	author = {Hawking, S. W.},
	journal = {Phys. Rev. D},
	volume = {14},
	issue = {10},
	pages = {2460--2473},
	numpages = {0},
	year = {1976},
	month = {Nov},
	publisher = {American Physical Society},
	doi = {10.1103/PhysRevD.14.2460},
	url = {https://link.aps.org/doi/10.1103/PhysRevD.14.2460}
}

@misc{byun2026quantumsimulationtraversablewormholeinspiredquantum,
	title={Quantum simulation of traversable-wormhole-inspired quantum teleportation in a chaotic binary sparse SYK model}, 
	author={Moongul Byun and Keun-Young Kim and Hyeonsoo Lee},
	year={2026},
	eprint={2604.10090},
	archivePrefix={arXiv},
	primaryClass={hep-th},
	url={https://arxiv.org/abs/2604.10090}, 
}

@article{10.1093/ptep/ptad147,
	author = {Nakayama, Yasuaki and Miyata, Akihiro and Ugajin, Tomonori},
	title = {The Petz (lite) recovery map for the scrambling channel},
	journal = {Progress of Theoretical and Experimental Physics},
	volume = {2023},
	number = {12},
	pages = {123B04},
	year = {2023},
	month = {12},
	issn = {2050-3911},
	doi = {10.1093/ptep/ptad147},
	url = {https://doi.org/10.1093/ptep/ptad147},
}

@article{Landsman2019,
	author = {Landsman, K. A. and Figgatt, C. and Schuster, T. and Linke, N. M. and Yoshida, B. and Yao, N. Y. and Monroe, C.},
	date = {2019/03/01},
	doi = {10.1038/s41586-019-0952-6},
	id = {Landsman2019},
	isbn = {1476-4687},
	journal = {Nature},
	number = {7746},
	pages = {61--65},
	title = {Verified quantum information scrambling},
	url = {https://doi.org/10.1038/s41586-019-0952-6},
	volume = {567},
	year = {2019}}

@article{PhysRevD.110.026010,
	title = {Quantum information recovery from a black hole with a projective measurement},
	author = {Li, Ran and Wang, Jin},
	journal = {Phys. Rev. D},
	volume = {110},
	issue = {2},
	pages = {026010},
	numpages = {20},
	year = {2024},
	month = {Jul},
	publisher = {American Physical Society},
	doi = {10.1103/PhysRevD.110.026010},
	url = {https://link.aps.org/doi/10.1103/PhysRevD.110.026010}
}

@article{PhysRevD.109.044005,
	title = {Information retrieval from Hawking radiation in the non-isometric model of black hole interior: Theory and quantum simulation},
	author = {Li, Ran and Wang, Xuanhua and Zhang, Kun and Wang, Jin},
	journal = {Phys. Rev. D},
	volume = {109},
	issue = {4},
	pages = {044005},
	numpages = {28},
	year = {2024},
	month = {Feb},
	publisher = {American Physical Society},
	doi = {10.1103/PhysRevD.109.044005},
	url = {https://link.aps.org/doi/10.1103/PhysRevD.109.044005}
}

@article{PhysRevResearch.7.023032,
	title = {Simulating Floquet scrambling circuits on trapped-ion quantum computers},
	author = {Seki, Kazuhiro and Kikuchi, Yuta and Hayata, Tomoya and Yunoki, Seiji},
	journal = {Phys. Rev. Res.},
	volume = {7},
	issue = {2},
	pages = {023032},
	numpages = {25},
	year = {2025},
	month = {Apr},
	publisher = {American Physical Society},
	doi = {10.1103/PhysRevResearch.7.023032},
	url = {https://link.aps.org/doi/10.1103/PhysRevResearch.7.023032}
}

@article{PhysRevX.11.021010,
	title = {Quantum Information Scrambling on a Superconducting Qutrit Processor},
	author = {Blok, M. S. and Ramasesh, V. V. and Schuster, T. and O'Brien, K. and Kreikebaum, J. M. and Dahlen, D. and Morvan, A. and Yoshida, B. and Yao, N. Y. and Siddiqi, I.},
	journal = {Phys. Rev. X},
	volume = {11},
	issue = {2},
	pages = {021010},
	numpages = {21},
	year = {2021},
	month = {Apr},
	publisher = {American Physical Society},
	doi = {10.1103/PhysRevX.11.021010},
	url = {https://link.aps.org/doi/10.1103/PhysRevX.11.021010}
}

@article{Kim2023,
	author = {Kim, MuSeong and Hwang, Mi-Ra and Jung, Eylee and Park, DaeKil},
	date = {2023/04/18},
	doi = {10.1007/s11128-023-03922-5},
	id = {Kim2023},
	isbn = {1573-1332},
	journal = {Quantum Information Processing},
	number = {4},
	pages = {176},
	title = {Scrambling and quantum teleportation},
	url = {https://doi.org/10.1007/s11128-023-03922-5},
	volume = {22},
	year = {2023}}

@article{tm83-sxpm,
	title = {Experimental simulation of postselected closed timelike curves for decoding scrambled quantum information},
	author = {Huang, Yi-Te and Huang, Hsiang-Wei and Lin, Jhen-Dong and Miranowicz, Adam and Lambert, Neill and Chen, Guang-Yin and Nori, Franco and Chen, Yueh-Nan},
	journal = {Phys. Rev. Res.},
	volume = {8},
	issue = {2},
	pages = {023084},
	numpages = {11},
	year = {2026},
	month = {Apr},
	publisher = {American Physical Society},
	doi = {10.1103/tm83-sxpm},
	url = {https://link.aps.org/doi/10.1103/tm83-sxpm}
}

@article{PhysRevLett.119.180509,
	title = {Error Mitigation for Short-Depth Quantum Circuits},
	author = {Temme, Kristan and Bravyi, Sergey and Gambetta, Jay M.},
	journal = {Phys. Rev. Lett.},
	volume = {119},
	issue = {18},
	pages = {180509},
	numpages = {5},
	year = {2017},
	month = {Nov},
	publisher = {American Physical Society},
	doi = {10.1103/PhysRevLett.119.180509},
	url = {https://link.aps.org/doi/10.1103/PhysRevLett.119.180509}
}

@article{PhysRevX.7.021050,
	title = {Efficient Variational Quantum Simulator Incorporating Active Error Minimization},
	author = {Li, Ying and Benjamin, Simon C.},
	journal = {Phys. Rev. X},
	volume = {7},
	issue = {2},
	pages = {021050},
	numpages = {14},
	year = {2017},
	month = {Jun},
	publisher = {American Physical Society},
	doi = {10.1103/PhysRevX.7.021050},
	url = {https://link.aps.org/doi/10.1103/PhysRevX.7.021050}
}

@article{PhysRevD.103.046004,
	title = {Observer-dependent black hole interior from operator collision},
	author = {Yoshida, Beni},
	journal = {Phys. Rev. D},
	volume = {103},
	issue = {4},
	pages = {046004},
	numpages = {24},
	year = {2021},
	month = {Feb},
	publisher = {American Physical Society},
	doi = {10.1103/PhysRevD.103.046004},
	url = {https://link.aps.org/doi/10.1103/PhysRevD.103.046004}
}

@article{PhysRevResearch.2.043024,
	title = {Realizing the Hayden-Preskill protocol with coupled Dicke models},
	author = {Cheng, Yanting and Liu, Chang and Guo, Jinkang and Chen, Yu and Zhang, Pengfei and Zhai, Hui},
	journal = {Phys. Rev. Res.},
	volume = {2},
	issue = {4},
	pages = {043024},
	numpages = {10},
	year = {2020},
	month = {Oct},
	publisher = {American Physical Society},
	doi = {10.1103/PhysRevResearch.2.043024},
	url = {https://link.aps.org/doi/10.1103/PhysRevResearch.2.043024}
}

@article{PhysRevResearch.6.L022021,
	title = {Hayden-Preskill recovery in Hamiltonian systems},
	author = {Nakata, Yoshifumi and Tezuka, Masaki},
	journal = {Phys. Rev. Res.},
	volume = {6},
	issue = {2},
	pages = {L022021},
	numpages = {8},
	year = {2024},
	month = {Apr},
	publisher = {American Physical Society},
	doi = {10.1103/PhysRevResearch.6.L022021},
	url = {https://link.aps.org/doi/10.1103/PhysRevResearch.6.L022021}
}

@article{PhysRevD.100.086001,
	title = {Soft mode and interior operator in the Hayden-Preskill thought experiment},
	author = {Yoshida, Beni},
	journal = {Phys. Rev. D},
	volume = {100},
	issue = {8},
	pages = {086001},
	numpages = {15},
	year = {2019},
	month = {Oct},
	publisher = {American Physical Society},
	doi = {10.1103/PhysRevD.100.086001},
	url = {https://link.aps.org/doi/10.1103/PhysRevD.100.086001}
}

@article{PhysRevResearch.2.043164,
	title = {Scrambling and decoding the charged quantum information},
	author = {Liu, Junyu},
	journal = {Phys. Rev. Res.},
	volume = {2},
	issue = {4},
	pages = {043164},
	numpages = {26},
	year = {2020},
	month = {Oct},
	publisher = {American Physical Society},
	doi = {10.1103/PhysRevResearch.2.043164},
	url = {https://link.aps.org/doi/10.1103/PhysRevResearch.2.043164}
}

@article{Nakata2023blackholesasclouded,
	doi = {10.22331/q-2023-02-21-928},
	url = {https://doi.org/10.22331/q-2023-02-21-928},
	title = {Black holes as clouded mirrors: the {H}ayden-{P}reskill protocol with symmetry},
	author = {Nakata, Yoshifumi and Wakakuwa, Eyuri and Koashi, Masato},
	journal = {{Quantum}},
	issn = {2521-327X},
	publisher = {{Verein zur F{\"{o}}rderung des Open Access Publizierens in den Quantenwissenschaften}},
	volume = {7},
	pages = {928},
	month = feb,
	year = {2023}
}

@misc{tajima2022universallimitationquantuminformation,
	title={Universal limitation of quantum information recovery: symmetry versus coherence}, 
	author={Hiroyasu Tajima and Keiji Saito},
	year={2022},
	eprint={2103.01876},
	archivePrefix={arXiv},
	primaryClass={quant-ph},
	url={https://arxiv.org/abs/2103.01876}, 
}

@misc{sun2026postselectionprobabilityfidelitybidirectional,
	title={Post-Selection Probability and Fidelity of Bidirectional Teleportation}, 
	author={Ning Sun and Lei Feng and Pengfei Zhang},
	year={2026},
	eprint={2606.17251},
	archivePrefix={arXiv},
	primaryClass={quant-ph},
	url={https://arxiv.org/abs/2606.17251}, 
}

@misc{vikram2026bidirectionalteleportationusingscrambling,
	title={Bidirectional teleportation using scrambling dynamics: a practical protocol}, 
	author={Amit Vikram and Edwin Chaparro and Muhammad Miskeen Khan and Andrew Lucas and Chris Akers and Ana Maria Rey},
	year={2026},
	eprint={2601.15536},
	archivePrefix={arXiv},
	primaryClass={quant-ph},
	url={https://arxiv.org/abs/2601.15536}, 
}

@article{PhysRevLett.127.270502,
	title = {Mitigating Depolarizing Noise on Quantum Computers with Noise-Estimation Circuits},
	author = {Urbanek, Miroslav and Nachman, Benjamin and Pascuzzi, Vincent R. and He, Andre and Bauer, Christian W. and de Jong, Wibe A.},
	journal = {Phys. Rev. Lett.},
	volume = {127},
	issue = {27},
	pages = {270502},
	numpages = {6},
	year = {2021},
	month = {Dec},
	publisher = {American Physical Society},
	doi = {10.1103/PhysRevLett.127.270502},
	url = {https://link.aps.org/doi/10.1103/PhysRevLett.127.270502}
}

@article{PhysRevB.98.014309,
	title = {Entanglement features of random Hamiltonian dynamics},
	author = {You, Yi-Zhuang and Gu, Yingfei},
	journal = {Phys. Rev. B},
	volume = {98},
	issue = {1},
	pages = {014309},
	numpages = {15},
	year = {2018},
	month = {Jul},
	publisher = {American Physical Society},
	doi = {10.1103/PhysRevB.98.014309},
	url = {https://link.aps.org/doi/10.1103/PhysRevB.98.014309}
}

@article{Preskill2018quantumcomputingin,
	doi = {10.22331/q-2018-08-06-79},
	url = {https://doi.org/10.22331/q-2018-08-06-79},
	title = {Quantum {C}omputing in the {NISQ} era and beyond},
	author = {Preskill, John},
	journal = {{Quantum}},
	issn = {2521-327X},
	publisher = {{Verein zur F{\"{o}}rderung des Open Access Publizierens in den Quantenwissenschaften}},
	volume = {2},
	pages = {79},
	month = aug,
	year = {2018}
}

@article{PhysRevD.97.066023,
	title = {Qubit transport model for unitary black hole evaporation without firewalls},
	author = {Osuga, Kento and Page, Don N.},
	journal = {Phys. Rev. D},
	volume = {97},
	issue = {6},
	pages = {066023},
	numpages = {7},
	year = {2018},
	month = {Mar},
	publisher = {American Physical Society},
	doi = {10.1103/PhysRevD.97.066023},
	url = {https://link.aps.org/doi/10.1103/PhysRevD.97.066023}
}

@article{PhysRevD.104.074518,
	title = {Diagnosis of information scrambling from Hamiltonian evolution under decoherence},
	author = {Hayata, Tomoya and Hidaka, Yoshimasa and Kikuchi, Yuta},
	journal = {Phys. Rev. D},
	volume = {104},
	issue = {7},
	pages = {074518},
	numpages = {25},
	year = {2021},
	month = {Oct},
	publisher = {American Physical Society},
	doi = {10.1103/PhysRevD.104.074518},
	url = {https://link.aps.org/doi/10.1103/PhysRevD.104.074518}
}

@article{Mao2026,
	author = {Mao, Weibo and Takayanagi, Tadashi},
	date = {2026/03/24},
	doi = {10.1007/JHEP03(2026)232},
	id = {Mao2026},
	isbn = {1029-8479},
	journal = {Journal of High Energy Physics},
	number = {3},
	pages = {232},
	title = {Hayden-Preskill model via local quenches},
	url = {https://doi.org/10.1007/JHEP03(2026)232},
	volume = {2026},
	year = {2026}}

@article{Penington2022,
	author = {Penington, Geoff and Shenker, Stephen H. and Stanford, Douglas and Yang, Zhenbin},
	date = {2022/03/30},
	doi = {10.1007/JHEP03(2022)205},
	id = {Penington2022},
	isbn = {1029-8479},
	journal = {Journal of High Energy Physics},
	number = {3},
	pages = {205},
	title = {Replica wormholes and the black hole interior},
	url = {https://doi.org/10.1007/JHEP03(2022)205},
	volume = {2022},
	year = {2022}}

@article{PhysRevLett.71.1291,
	title = {Average entropy of a subsystem},
	author = {Page, Don N.},
	journal = {Phys. Rev. Lett.},
	volume = {71},
	issue = {9},
	pages = {1291--1294},
	numpages = {0},
	year = {1993},
	month = {Aug},
	publisher = {American Physical Society},
	doi = {10.1103/PhysRevLett.71.1291},
	url = {https://link.aps.org/doi/10.1103/PhysRevLett.71.1291}
}

@article{doi:10.1142/S1230161208000043,
	author = {Hayden, Patrick and Horodecki, Micha\l{} and Winter, Andreas and Yard, Jon},
	title = {A Decoupling Approach to the Quantum Capacity},
	journal = {Open Systems \& Information Dynamics},
	volume = {15},
	number = {01},
	pages = {7-19},
	year = {2008},
	doi = {10.1142/S1230161208000043},
	URL = { 
	https://doi.org/10.1142/S1230161208000043
	},
	eprint = { 
	https://doi.org/10.1142/S1230161208000043
	}
}

@inproceedings{10.1145/237814.237866,
	author = {Grover, Lov K.},
	title = {A fast quantum mechanical algorithm for database search},
	year = {1996},
	isbn = {0897917855},
	publisher = {Association for Computing Machinery},
	address = {New York, NY, USA},
	url = {https://doi.org/10.1145/237814.237866},
	doi = {10.1145/237814.237866},
	booktitle = {Proceedings of the Twenty-Eighth Annual ACM Symposium on Theory of Computing},
	pages = {212–219},
	numpages = {8},
	location = {Philadelphia, Pennsylvania, USA},
	series = {STOC '96}
}

@article{PhysRevA.107.032418,
	title = {Scrambling and quantum chaos indicators from long-time properties of operator distributions},
	author = {Omanakuttan, Sivaprasad and Chinni, Karthik and Blocher, Philip Daniel and Poggi, Pablo M.},
	journal = {Phys. Rev. A},
	volume = {107},
	issue = {3},
	pages = {032418},
	numpages = {15},
	year = {2023},
	month = {Mar},
	publisher = {American Physical Society},
	doi = {10.1103/PhysRevA.107.032418},
	url = {https://link.aps.org/doi/10.1103/PhysRevA.107.032418}
}
		
	\end{document}